\documentclass[12pt]{article}
\usepackage[utf8]{inputenc}
\usepackage[T1]{fontenc}
\usepackage[margin=1in]{geometry}
\usepackage{amsmath}
\usepackage{graphicx,psfrag,epsf}
\usepackage{enumerate}
\usepackage{natbib}
\usepackage{url}
\usepackage{multicol}
\usepackage{tikz}
\usetikzlibrary{shapes.geometric, arrows}
\usepackage{array}
\usepackage{hyperref}
\hypersetup{
    colorlinks=true,
    citecolor=black,
    linkcolor=black,
    filecolor=black,      
    urlcolor=black}

\usepackage{enumitem}
\usepackage{amsfonts}
\usepackage{amssymb}
\usepackage{booktabs}

\def\bSig\mathbf{\Sigma}

\newcommand{\indep}{\rotatebox[origin=c]{90}{$\models$}}

\begin{document}

\title{\bf Dealing with multiple intercurrent events using hypothetical and treatment policy strategies simultaneously}

\author{Camila Olarte Parra \\
Unit of Epidemiology, Institute of Enviromental Medicine, \\ Karolinska Universitet, Sweden \vspace{0.3cm}\\
    Rhian M. Daniel \\
    Division of Population Medicine, \\
    Cardiff University, UK \\
\vspace{0.3cm}\\
    Jonathan W. Bartlett \\
    Department of Medical Statistics, \\ 
    London School of Hygiene and Tropical Medicine, UK}

\date{}

\maketitle



\begin{abstract}
{To precisely define the treatment effect of interest in a clinical trial, the ICH E9 estimand addendum describes that relevant so-called intercurrent events should be identified and strategies specified to deal with them. Handling intercurrent events with different strategies leads to different estimands. In this paper, we focus on estimands that involve addressing one intercurrent event with the treatment policy strategy and another with the hypothetical strategy. We define these estimands using potential outcomes and causal diagrams, considering the possible causal relationships between the two intercurrent events and other variables. We show that there are different causal estimand definitions and assumptions one could adopt, each having different implications for estimation, which is demonstrated in a simulation study. The different considerations are illustrated conceptually using a diabetes trial as an example.}

    
\end{abstract}

\maketitle



\newpage
\section{Introduction}\label{sec:intro}

The ICH E9 addendum on estimands introduces the notion of `intercurrent event' (ICE) to refer to events that occur after treatment initiation and that can either prevent the occurrence of the outcome or affect its interpretation \citep{ICHE9Addendum}. Examples of intercurrent events include treatment discontinuation, addition of rescue medication and death prior to outcome assessment. The addendum indicates that when designing clinical trials, relevant ICEs should be identified and strategies to handle them should be chosen. Identifying ICEs and strategies to address them is part of defining the target estimand of a trial \citep{Mallinckrodt2020Estimands}. The estimand precisely defines the treatment effect of interest. In particular, dealing with ICEs through different strategies leads to different estimands \citep{Lipkovich2020Estimands}. The addendum outlines different such strategies in words but does not give a mathematical characterisation of them, and moreover it does not discuss statistical estimation. 

In this paper we explore from a causal inference perspective the definition, meaning, identification and estimation of estimands that involve both the treatment policy and hypothetical strategies, building on earlier work \citep{Lipkovich2020Estimands,Olarte2023hypothetical,Ocampo2022SWIG}. We start by describing a motivating example in diabetes and discussing potential ICEs of interest that are to be addressed with these strategies (Section \ref{sec:example}). We then discuss the different estimands that result from handling different ICEs with either strategy or a combination of both, defining them using the language of potential outcomes and causal diagrams (Section \ref{sec:estimands}). We describe identification assumptions for these different estimands starting with a setting where the ICEs can only occur at single time point (Section \ref{sec:identifiability}). We then extend the results to a longitudinal setting where the ICEs can happen at multiple time points and discuss the implications of these for the statistical analysis in terms of which variables should be adjusted for and how (Section \ref{sec:long}). As a proof of concept and illustration of the estimation process, we conducted a simulation study under the different data generating mechanisms discussed (Section \ref{sec:simulations}). Finally, we  re-consider the motivating example in light of our results (Section \ref{sec:revisiting}) and close with discussion and recommendations (Section \ref{sec:conc}). 

\section{Motivating example}
\label{sec:example}
As a motivating example we consider a randomised controlled trial (RCT) where type 2 diabetes patients on metformin monotherapy experiencing inadequate glycaemic control were randomised at baseline to receive add-on dapagliflozin, dapagliflozin and saxagliptin, or glimepiride, the results of which were reported by Muller \textit{et al} \citep{Muller2018Diabetes}.  In the trial, patients' HbA1c and fasting plasma glucose (FGP) were measured periodically to assess their response, with the final assessment made after 52 weeks of follow-up. Insulin was indicated as rescue medication for patients with inadequate glycaemic control in the trial. The study protocol specified the threshold of FPG, for earlier visits, and of HbA1c, at the last two visits, which if exceeded, would lead to considering initiation of rescue medication.

In such a trial, rescue medication should be available for medical and ethical reasons, but its use can mask the effect of the new drug as compared to the control \citep{Holzhauer2015DiabetesRescueMed}. Thus, interest may lie in estimating the treatment effect under the hypothetical scenario where rescue medication would not have been made available, so that the resulting treatment effect is not affected by any potential benefit from rescue drugs. This is in line with the latest recommendation from the European Medicines Agency (EMA) of handling additional (rescue) medication with the hypothetical strategy in the setting of diabetes trials \citep{EMAdiabetes}.

During follow-up, some patients discontinued their randomised medication for different reasons including lack of efficacy or adverse events. As this would be expected to happen in clinical practice, we could consider dealing with treatment discontinuation using the treatment policy strategy, which means that our interest is in the outcomes as they would be observed regardless of whether randomised treatment was discontinued or not. The EMA recommends this strategy for treatment discontinuation in diabetes trials on the basis of there being no benefit after treatment discontinuation \citep{EMAdiabetes}. However, in many disease areas, in both clinical trials and in real clinical practice, patients who discontinue their randomised (or initial) treatment will then begin taking an alternative treatment such that the treatment policy strategy would include any effects of this subsequent treatment. In light of this, we could also consider dealing with treatment discontinuation with the hypothetical strategy, where we target the treatment effect in the hypothetical scenario where patients who would ordinarily discontinue treatment would have instead somehow been made to continue receiving their assigned treatment. This could potentially be informative for regulators and patients of the expected benefit if assigned treatment could somehow be adhered to by all patients. Precise definition and estimation of such an estimand can however be problematic, as we discuss further in Section \ref{sec:identifiability}. 

\section{Estimand definitions}
\label{sec:estimands}
In this section we define different possible estimands in the context of a simplified trial setup with two ICEs which can each occur (or not) at one follow-up time point, with the final outcome measured after this. To do this, we will introduce some notation. Let $A$ denote the treatment randomly assigned at baseline, $R$ the use of rescue medication, $D$ treatment discontinuation and $Y$ the outcome of interest. We moreover use potential outcome notation, so that $Y^a$ is the potential outcome when we set treatment to level $a$ \citep{hernan2004definition}. As other (baseline) variables are not required to define the estimands, we will omit them for now and introduce them in Section \ref{sec:identifiability}.

\begin{figure}
    Possible causal structures
    \newline
    \begin{minipage}{0.30\textwidth}
    
    a) \resizebox{0.9\textwidth}{!}{\begin{tikzpicture}
    \node (A) at (0.00,1.00) {$A$};
    \node (D) at (3.00,2.00) {$D$};
    \node (R) at (3.00,0.00) {$R$};
    \node (Y) at (6.00,1.00) {$Y$};
    
    \draw [->] (A) edge (D);
    \draw [->] (A) edge (R);
    \draw [->] (A) edge (Y);

    \draw [->] (D) edge (Y);
    \draw [->] (R) edge (Y);
    
    \end{tikzpicture}}
    \end{minipage}
    \hspace{0.02\textwidth}
    \begin{minipage}{0.30\textwidth}
    b) \resizebox{0.9\textwidth}{!}{\begin{tikzpicture}
    \node (A) at (0.00,1.00) {$A$};
    \node (D) at (4.00,2.00) {$D$};
    \node (R) at (2.00,0.00) {$R$};
    \node (Y) at (6.00,1.00) {$Y$};
    
    \draw [->] (A) edge (R);
    \draw [->] (A) edge (D);
    \draw [->] (A) edge (Y);

    \draw [->] (R) edge (Y);
    \draw [->] (R) edge (D);
    \draw [->] (D) edge (Y);
    
    \end{tikzpicture}} 
    
    \end{minipage}
    \hspace{0.02\textwidth}
    \begin{minipage}{0.30\textwidth}
    c) \resizebox{0.9\textwidth}{!}{\begin{tikzpicture}
\node (A) at (0.00,1.00) {$A$};
\node (D) at (2.00,2.00) {$D$};
\node (R) at (4.00,0.00) {$R$};
\node (Y) at (6.00,1.00) {$Y$};

\draw [->] (A) edge (R);
\draw [->] (A) edge (D);
\draw [->] (A) edge (Y);

\draw [->] (R) edge (Y);
\draw [->] (D) edge (R);
\draw [->] (D) edge (Y);

\end{tikzpicture}}
    
    \end{minipage}\vspace{0.5cm}
    
    Handling both ICEs with treatment policy strategy
    \newline
    \begin{minipage}{0.30\textwidth}

    d)
    \resizebox{0.9\textwidth}{!}{\begin{tikzpicture}
    \node (A) at (0.00,1.00) {$A$};
    \node[rectangle,draw] (a) at (0.50,1.00) {$a$};
    \node (D) at (3.00,2.00) {$D^a$};
    \node (R) at (3.00,0.00) {$R^a$};
    \node (Y) at (6.00,1.00) {$Y^{a}$};
    
    \draw [->] (a) edge (D);
    \draw [->] (a) edge (R);
    \draw [->] (a) edge (Y);

    \draw [->] (D) edge (Y);
    \draw [->] (R) edge (Y);
    
    \end{tikzpicture}}    
    \end{minipage}
    \hspace{0.02\textwidth}
    \begin{minipage}{0.30\textwidth}
    e)
    \resizebox{0.9\textwidth}{!}{\begin{tikzpicture}
    \node (A) at (0.00,1.00) {$A$};
    \node[rectangle,draw] (a) at (0.50,1.00) {$a$};
    \node (D) at (4.00,2.00) {$D^{a}$};
    \node (R) at (2.00,0.00) {$R^a$};
    \node (Y) at (6.00,1.00) {$Y^{a}$};
    
    \draw [->] (a) edge (D);
    \draw [->] (a) edge (R);
    \draw [->] (a) edge (Y);

    \draw [->] (D) edge (Y);
    \draw [->] (R) edge (D);
    \draw [->] (R) edge (Y);
    
    \end{tikzpicture}}
    \end{minipage}
    \hspace{0.02\textwidth}
    \begin{minipage}{0.30\textwidth}
    f)
    \resizebox{0.9\textwidth}{!}{\begin{tikzpicture}
    \node (A) at (0.00,1.00) {$A$};
    \node[rectangle,draw] (a) at (0.50,1.00) {$a$};
    \node (D) at (2.00,2.00) {$D^{a}$};
    \node (R) at (4.00,0.00) {$R^a$};
    \node (Y) at (6.50,1.00) {$Y^{a}$};
    
    \draw [->] (a) edge (D);
    \draw [->] (a) edge (R);
    \draw [->] (a) edge (Y);

    \draw [->] (D) edge (Y);
    \draw [->] (D) edge (R);
    \draw [->] (R) edge (Y);

    \end{tikzpicture}} 
    \end{minipage}\vspace{0.5cm}
    
    Handling both ICEs with hypothetical strategy

    \begin{minipage}{0.30\textwidth}
    g)
    \resizebox{0.9\textwidth}{!}{\begin{tikzpicture}
    \node (A) at (0.00,1.00) {$A$};
    \node[rectangle,draw] (a) at (0.50,1.00) {$a$};
    \node (D) at (3.00,2.00) {$D^a$};
    \node[rectangle,draw] (d) at (4.00,2.00) {$d=0$};
    \node (R) at (3.00,0.00) {$R^a$};
    \node[rectangle,draw] (r) at (4.00,0.00) {$r=0$};
    \node (Y) at (6.00,1.00) {$Y^{a,d=0,r=0}$};
    
    \draw [->] (a) edge (D);
    \draw [->] (a) edge (R);
    \draw [->] (a) edge (Y);

    \draw [->] (d) edge (Y);
    \draw [->] (r) edge (Y);
    
    \end{tikzpicture}}
    \end{minipage}   \hspace{0.02\textwidth}
    \begin{minipage}{0.30\textwidth}
    h)
     \resizebox{0.9\textwidth}{!}{\begin{tikzpicture}
    \node (A) at (0.00,1.00) {$A$};
    \node[rectangle,draw] (a) at (0.50,1.00) {$a$};
    \node (D) at (3.80,2.00) {$D^{a,r=0}$};
    \node[rectangle,draw] (d) at (5.20,2.00) {$d=0$};
    \node (R) at (2.00,0.00) {$R^a$};
    \node[rectangle,draw] (r) at (3.00,0.00) {$r=0$};
    \node (Y) at (6.00,1.00) {$Y^{a,d=0,r=0}$};
    
    \draw [->] (a) edge (D);
    \draw [->] (a) edge (R);
    \draw [->] (a) edge (Y);

    \draw [->] (d) edge (Y);
    \draw [->] (r) edge (D);
    \draw [->] (r) edge (Y);
    
    \end{tikzpicture}}
    \end{minipage}
    \hspace{0.02\textwidth}
    \begin{minipage}{0.30\textwidth}
    i)
    \resizebox{0.9\textwidth}{!}{\begin{tikzpicture}
    \node (A) at (0.00,1.00) {$A$};
    \node[rectangle,draw] (a) at (0.50,1.00) {$a$};
    \node (D) at (2.00,2.00) {$D^{a}$};
    \node[rectangle,draw] (d) at (3.00,2.00) {$d=0$};
    \node (R) at (3.80,0.00) {$R^{a,d=0}$};
    \node[rectangle,draw] (r) at (5.20,0.00) {$r=0$};
    \node (Y) at (6.50,1.00) {$Y^{a,d=0,r=0}$};
    
    \draw [->] (a) edge (D);
    \draw [->] (a) edge (R);
    \draw [->] (a) edge (Y);

    \draw [->] (d) edge (Y);
    \draw [->] (d) edge (R);
    \draw [->] (r) edge (Y);

    \end{tikzpicture}}
    \end{minipage}\vspace{0.5cm}
    
    Handling ICE $D$ with treatment policy and ICE $R$ with hypothetical strategy
    
    \begin{minipage}{0.30\textwidth}
    j)
    \resizebox{0.9\textwidth}{!}{\begin{tikzpicture}
    \node (A) at (0.00,1.00) {$A$};
    \node[rectangle,draw] (a) at (0.50,1.00) {$a$};
    \node (D) at (3.00,2.00) {$D^a$};
    \node (R) at (3.00,0.00) {$R^a$};
    \node[rectangle,draw] (r) at (4.00,0.00) {$r=0$};
    \node (Y) at (6.00,1.00) {$Y^{a,r=0}$};
    
    \draw [->] (a) edge (D);
    \draw [->] (a) edge (R);
    \draw [->] (a) edge (Y);

    \draw [->] (D) edge (Y);
    \draw [->] (r) edge (Y);
    
    \end{tikzpicture}}
    \end{minipage}
    \hspace{0.02\textwidth}    
    \begin{minipage}{0.30\textwidth}
    k)
    \resizebox{0.9\textwidth}{!}{\begin{tikzpicture}
    \node (A) at (0.00,1.00) {$A$};
    \node[rectangle,draw] (a) at (0.50,1.00) {$a$};
    \node (D) at (4.00,2.00) {$D^{a,r=0}$};
    \node (R) at (2.00,0.00) {$R^a$};
    \node[rectangle,draw] (r) at (3.00,0.00) {$r=0$};
    \node (Y) at (6.00,1.00) {$Y^{a,r=0}$};
    
    \draw [->] (a) edge (D);
    \draw [->] (a) edge (R);
    \draw [->] (a) edge (Y);

    \draw [->] (D) edge (Y);
    \draw [->] (r) edge (D);
    \draw [->] (r) edge (Y);
    
    \end{tikzpicture}}
    \end{minipage}
    \hspace{0.02\textwidth}
    \begin{minipage}{0.30\textwidth}
    l)
    \resizebox{0.9\textwidth}{!}{\begin{tikzpicture}
    \node (A) at (0.00,1.00) {$A$};
    \node[rectangle,draw] (a) at (0.50,1.00) {$a$};
    \node (D) at (2.00,2.00) {$D^{a}$};
    \node (R) at (4.00,0.00) {$R^a$};
    \node[rectangle,draw] (r) at (5.00,0.00) {$r=0$};
    \node (Y) at (6.50,1.00) {$Y^{a,r=0}$};
    
    \draw [->] (a) edge (D);
    \draw [->] (a) edge (R);
    \draw [->] (a) edge (Y);

    \draw [->] (D) edge (Y);
    \draw [->] (D) edge (R);
    \draw [->] (r) edge (Y);

    \end{tikzpicture}} 
    \end{minipage}\vspace{0.5cm}

    \caption{Simplified DAGs (first row) and SWIGs (later rows) to illustrate potential estimands of interest under different causal structures (columns) relating randomised treatment $A$, treatment discontinuation $D$, use of rescue medication $R$ and outcome $Y$. Each row represents the resulting estimand when handling the ICEs with different strategies.}
    \label{fig:estimands}
\end{figure}
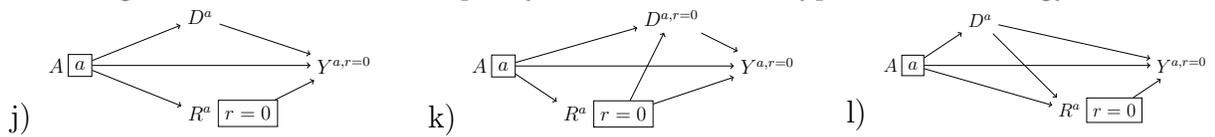

Figure \ref{fig:estimands} summarises possible causal structures for $A$, $R$, $D$ and $Y$ using directed acyclic graphs (DAGs) and single world intervention graphs (SWIGs) \citep{Ocampo2022SWIG}. Suppose that whether each ICE occurs (or not) happens at the same time (relative to baseline) for each individual, there could be three possible situations, represented by each column in the figure: 1) ICE $R$ and ICE $D$ do not affect each other, 2) ICE $R$ has an effect on ICE $D$ or 3) ICE $D$ has an effect on ICE $R$. Which structure is reasonable in a given setting can be considered given the time ordering between the two ICEs, although we acknowledge that in practice it may not be entirely clear that one ICE precedes another chronologically, or that this ordering is necessarily the same for all individuals. Given these three different DAGs, we can then consider the resulting estimands of handling both ICEs with the treatment policy strategy, both with the hypothetical strategy and, finally, one ICE with treatment policy and the other one with the hypothetical strategy, depicted in each row of Figure \ref{fig:estimands}.

To represent the estimands that result from handling the ICEs with different strategies, we make use of SWIGs \citep{Ocampo2022SWIG}. Each SWIG in Figure \ref{fig:estimands} is based on the DAG on the top of the corresponding column, and depicts the causal relationships when intervening to set one or more variables to a particular value. This is shown by splitting the node of the intervened variable to represent both the natural (in the world without intervention on the ICE) value of that variable and the value to which the variable is set to. The box around the set value in the split node represents that this is an intervened variable. The subsequent variables affected by the intervened node are then changed to their potential outcome under that value. The appearance of the relevant potential outcomes in the SWIGs (which are absent from DAGs) allows us to reason graphically about certain (conditional) exchangeability assumptions, which are required to identify the effect, as described in Section \ref{sec:identifiability}. 

When handling both ICEs with a treatment policy strategy, our estimand of interest is:
\begin{equation}
\label{eq:itt_estimand}
    E\left(Y^{a=1,D^{a},R^{a}}-Y^{a=0,D^{a},R^{a}}\right) = E\left(Y^{a=1}-Y^{a=0}\right)
\end{equation}
Here we are targeting the treatment effect including the effects of the occurrence of either ICE on the final outcome. Thus for this estimand, in each arm we have a combination of participants who received their assigned treatment throughout, those who received the treatment throughout and used rescue medication, those who discontinued randomised treatment and those who received rescue medication and discontinued randomised treatment. As there are no special (causal) considerations for the meaning or estimation of this estimand, save for the usual (often non-trivial) issue of missing data, we will not discuss it further.

Alternatively, handling both ICEs with the hypothetical strategy leads to the following estimand:
\begin{equation}
\label{eq:hypothetical_estimand}
    E\left(Y^{a=1,d=0,r=0}-Y^{a=0,d=0,r=0}\right)
\end{equation}
This is the treatment effect when we intervene to prevent both ICEs from occurring. In this case, we are considering a hypothetical scenario where we somehow enforce that all participants receive their assigned treatment throughout and do not receive rescue medication. 

If we use the treatment policy strategy for ICE $D$ and the hypothetical strategy for ICE $R$, as suggested by EMA diabetes guidelines, the resulting estimand is either \\ $E\left(Y^{a=1,D^{a},r=0}-Y^{a=0,D^{a},r=0}\right)$, as in Figure \ref{fig:estimands} j) and l) or $E\left(Y^{a=1,D^{a,r=0},r=0}-Y^{a=0,D^{a,r=0},r=0}\right)$, as in Figure \ref{fig:estimands} k). As we are not intervening on $D$, both estimands reduce to:
\begin{equation}
\label{eq:combined_estimand1}
    E\left(Y^{a=1,r=0}-Y^{a=0,r=0}\right)
\end{equation}
For brevity, we refer to this estimand as \emph{hypothetical estimand} throughout the paper. Here, we consider the treatment effect under the hypothetical scenario in which we intervene to prevent the ICE $R$ and let the ICE $D$ take its natural value under the assigned treatment $a$, or in the situation in column 2, where $R$ affects $D$, its value under treatment $a$ and setting $R$ to $0$. Thus we are interested in the treatment effect in a hypothetical trial where we have a combination of patients who continue their treatment throughout and patients who discontinue their randomised treatment but none who received rescue medication. 

Under Figure \ref{fig:estimands} k), in the real trial, there may be patients who did not discontinue because they received rescue medication who would have discontinued had rescue not been available. Thus in this situation, although we adopt the treatment policy strategy for $D$, the ICE $D$ takes the value it would take if rescue treatment were not permitted. That is, for this estimand, $D$ does not take its natural value for all individuals, but its value under the hypothetical situation where rescue is withheld. Given that lack of efficacy can lead to discontinuation, we might expect that withholding rescue medication would lead to an increase in treatment discontinuation. 
    
An alternative interpretation of what it means to use the hypothetical strategy for $R$ and the treatment policy strategy for $D$ is the scenario in which we intervene to prevent the ICE $R$ but let the ICE $D$ take its natural value under the assigned treatment $a$ and the value that $R$ would have taken under treatment $a$ (had $R$ not been intervened on). For the causal structures depicted in Figure \ref{fig:estimands} j) and l), the estimand is exactly the same as before (hypothetical estimand, \ref{eq:combined_estimand1}) because ICE $R$ does not affect ICE $D$. However for the situation in Figure \ref{fig:estimands} b), this estimand under this alternative interpretation is as follows:

\begin{equation}
\label{EQ:COMBINED_ESTIMAND2}
    E\left(Y^{a=1,r=0,D^{a=1,R^{a=1}}}-Y^{a=0,r=0,D^{a=0,R^{a=0}}}\right)
\end{equation}

This estimand, referred throughout the paper as \emph{cross-world hypothetical}, is difficult to understand since it is impossible to design even a hypothetical trial corresponding to it, because it involves at the same time setting $R$ to zero but letting $D$ take its natural value had $R$ not been intervened on. Put another way, the potential outcomes in this estimand do not correspond to a single intervention on a subset of variables in the DAG and therefore cannot be depicted in a SWIG.

\section{Identifiability}
\label{sec:identifiability}

In this section, we describe assumptions under which the different estimands discussed in Section \ref{sec:estimands} can be identified and estimated. To do that, we will apply the usual identifiability assumptions for time-varying treatments \citep{Hernan2020Ch19}. To discuss the identifiability assumptions, we will introduce further notation. As part of the trial, we assume that we will record baseline covariates ($L_0$) and post-baseline covariates ($L_1$) measured at a follow-up visit that precedes the ICE variables. Note that $L_0$ and $L_1$ may include baseline and intermediate measurements of the outcome respectively.

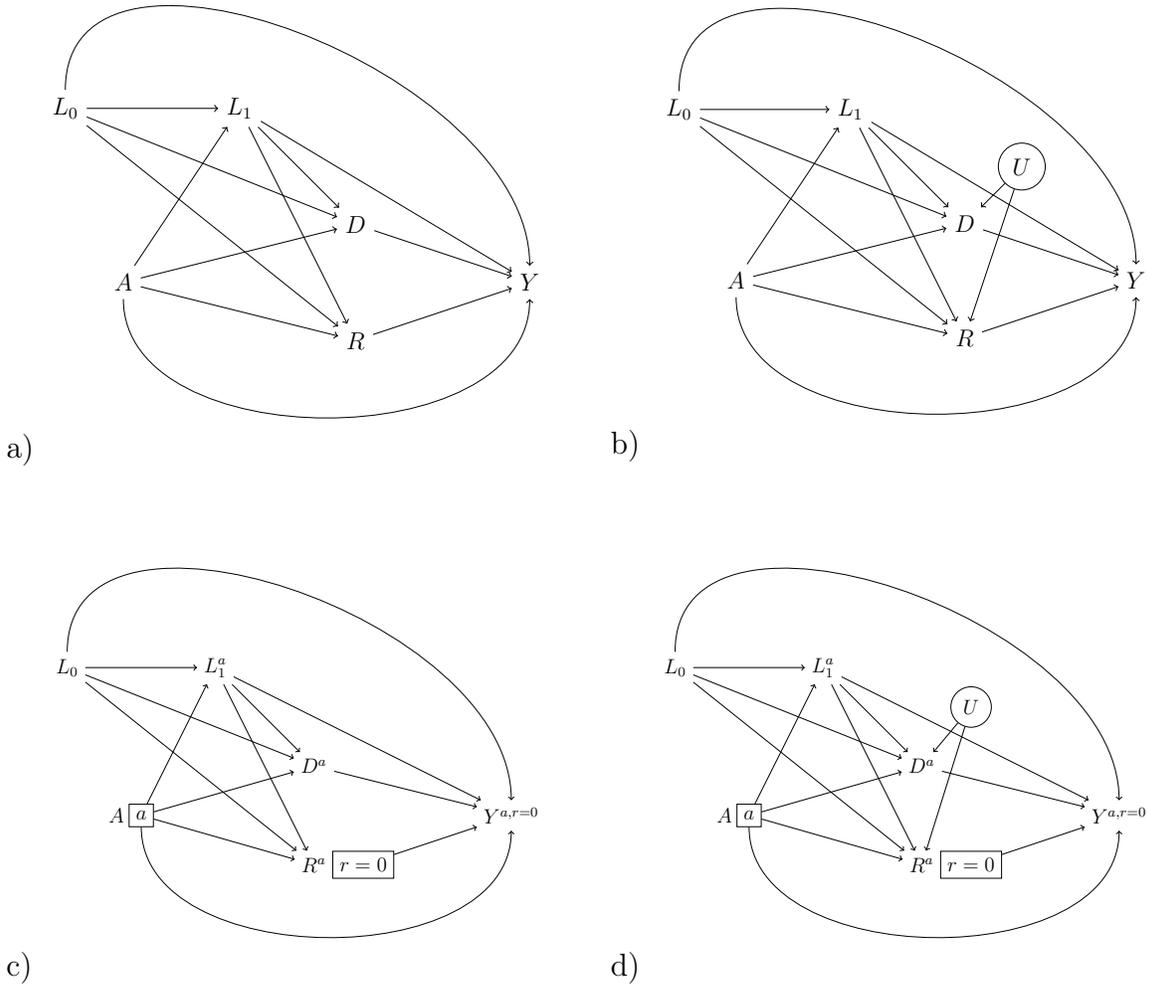
\begin{figure}
    \centering
        \begin{minipage}{0.45\textwidth}
    
    a) \resizebox{0.9\textwidth}{!}{\begin{tikzpicture}
    \node (A) at (1.00,1.00) {$A$};
    \node (M) at (5.00,2.00) {$D$};
    \node (H) at (5.00,0.00) {$R$};
    \node (L_0) at (0.00,4.00) {$L_0$};
    \node (L_1) at (3.00,4.00) {$L_1$};
    \node (Y) at (8.00,1.00) {$Y$};
    
    \draw [->] (L_0) edge (L_1);
    \draw [->] (L_0) edge (M);
    \draw [->] (L_0) edge (H);
    \draw [->] (L_0) to[out=90,in=90] (Y);
    
    \draw [->] (A) edge (M);
    \draw [->] (A) edge (H);
    \draw [->] (A) to[out=-90,in=-90] (Y);
    \draw [->] (A) edge (L_1);
    
    \draw [->] (M) edge (Y);
    \draw [->] (H) edge (Y);
    
    \draw [->] (L_1) edge (M);
    \draw [->] (L_1) edge (H);
    \draw [->] (L_1) edge (Y);
    
    \end{tikzpicture}}
    \end{minipage}
    \hspace{0.02\textwidth}
    \begin{minipage}{0.45\textwidth}
    
    b) \resizebox{0.9\textwidth}{!}{
    \begin{tikzpicture}
    \node (A) at (1.00,1.00) {$A$};
    \node (M) at (5.00,2.00) {$D$};
    \node (H) at (5.00,0.00) {$R$};
    \node (L_0) at (0.00,4.00) {$L_0$};
    \node (L_1) at (3.00,4.00) {$L_1$};
    \node (Y) at (8.00,1.00) {$Y$};
    \node[circle,draw] (U) at (6.00,3.00) {$U$};
    
    \draw [->] (L_0) edge (L_1);
    \draw [->] (L_0) edge (M);
    \draw [->] (L_0) edge (H);
    \draw [->] (L_0) to[out=90,in=90] (Y);
    
    \draw [->] (A) edge (M);
    \draw [->] (A) edge (H);
    \draw [->] (A) to[out=-90,in=-90] (Y);
    \draw [->] (A) edge (L_1);
    
    \draw [->] (M) edge (Y);
    \draw [->] (H) edge (Y);
    
    \draw [->] (U) edge (M);
    \draw [->] (U) edge (H);
    
    \draw [->] (L_1) edge (M);
    \draw [->] (L_1) edge (H);
    \draw [->] (L_1) edge (Y);
    
    \end{tikzpicture}}
    \end{minipage}
    \begin{minipage}{0.45\textwidth}
    c) \resizebox{0.9\textwidth}{!}{
    \begin{tikzpicture}
    \node (A) at (1.00,1.00) {$A$};
    \node[rectangle,draw] (a) at (1.50,1.00) {$a$};
    \node (M) at (5.00,2.00) {$D^a$};
    \node (H) at (5.00,0.00) {$R^a$};
    \node[rectangle,draw] (h) at (6.00,0.00) {$r=0$};
    \node (L_0) at (0.00,4.00) {$L_0$};
    \node (L_1) at (3.00,4.00) {$L^a_1$};
    \node (Y) at (9.00,1.00) {$Y^{a,r=0}$};
    
    \draw [->] (L_0) edge (L_1);
    \draw [->] (L_0) edge (M);
    \draw [->] (L_0) edge (H);
    \draw [->] (L_0) to[out=90,in=90] (Y);
    
    \draw [->] (a) edge (M);
    \draw [->] (a) edge (H);
    \draw [->] (a) to[out=-90,in=-90] (Y);
    \draw [->] (a) edge (L_1);
    
    \draw [->] (M) edge (Y);
    \draw [->] (h) edge (Y);
    
    \draw [->] (L_1) edge (M);
    \draw [->] (L_1) edge (H);
    \draw [->] (L_1) edge (Y);
    
    \end{tikzpicture}}
    \end{minipage}
    \hspace{0.02\textwidth}
    \begin{minipage}{0.45\textwidth}
    d) \resizebox{0.9\textwidth}{!}{
    \begin{tikzpicture}
    \node (A) at (1.00,1.00) {$A$};
    \node[rectangle,draw] (a) at (1.50,1.00) {$a$};
    \node (M) at (5.00,2.00) {$D^a$};
    \node (H) at (5.00,0.00) {$R^a$};
    \node[rectangle,draw] (h) at (6.00,0.00) {$r=0$};
    \node (L_0) at (0.00,4.00) {$L_0$};
    \node (L_1) at (3.00,4.00) {$L^a_1$};
    \node (Y) at (9.00,1.00) {$Y^{a,r=0}$};
    \node[circle,draw] (U) at (6.00,3.20) {$U$};
    
    \draw [->] (L_0) edge (L_1);
    \draw [->] (L_0) edge (M);
    \draw [->] (L_0) edge (H);
    \draw [->] (L_0) to[out=90,in=90] (Y);
    
    \draw [->] (a) edge (M);
    \draw [->] (a) edge (H);
    \draw [->] (a) to[out=-90,in=-90] (Y);
    \draw [->] (a) edge (L_1);
    
    \draw [->] (M) edge (Y);
    \draw [->] (h) edge (Y);
    
    \draw [->] (U) edge (M);
    \draw [->] (U) edge (H);
    
    \draw [->] (L_1) edge (M);
    \draw [->] (L_1) edge (H);
    \draw [->] (L_1) edge (Y);
    
    \end{tikzpicture}}
    \end{minipage}
    \caption{Graphical representation using DAGs (a,b) and SWIGs (c,d) of a trial when setting $A$ to $a$ and $R$ to 0, where treatment $A$ is assigned at random, ICE $D$ handled by treatment policy and ICE $R$ is handled by hypothetical strategy. ICE $D$ and ICE $R$ do not have a causal link between them but both ICE may have an effect on outcome $Y$. $L_0$, $L_1$ and $U$ denote baseline variables, variables measured at $t=1$ and unmeasured variables affecting $D$ and $U$, respectively.}
    \label{fig:simultaneous_ICE}
\end{figure}

Figure \ref{fig:simultaneous_ICE} a) depicts a situation where the ICEs do not affect each other, similar to Figure \ref{fig:estimands} a), but now including $L_0$ and $L_1$. As treatment is assigned at random there are no arrows from $L_0$ to $A$. Note that we assume $L_1$ precedes $D$ and $R$ because $L_1$ includes any signs, symptoms or diagnostic tests taken at the follow-up visit that could inform the decision of initiating rescue medication and/or discontinuing from randomised treatment. 

To identify the causal effect of $A$ on $Y$ with the treatment policy strategy used for both ICEs, we assume that the observed outcome $Y$ corresponds to the potential outcome under the assigned treatment, i.e. that $Y^a=Y$ when $A=a$ (\emph{consistency assumption}). Also, we assume that the assigned treatment is independent of the potential outcome, which is the case when $A$ is assigned at random, making $Y^a \indep A$ (\emph{exchangeability assumption}). Finally, everyone should have a positive probability of receiving each treatment, which also holds in a randomised controlled trial (RCT) where usually $P(A=1)=P(A=0)=0.5$ (\emph{positivity assumption}). 

However, when we choose to handle either ICE with the hypothetical strategy, we need to extend these conditions to the corresponding hypothetical ICE as well. Figure \ref{fig:simultaneous_ICE} c) takes Figure \ref{fig:simultaneous_ICE} a) as input when we decide to handle $D$ with treatment policy and $R$ with the hypothetical strategy. Here, the consistency assumption means assuming that $R^a=R$ when $A=a$ and $Y^{a,r=0}=Y$ when $A=a$ and $R=0$, which in words means that for patients in the actual trial that did not receive rescue, their realised outcome equals the outcome they would have experienced in the hypothetical trial where rescue treatment is withheld from all patients. 

Following Figure \ref{fig:simultaneous_ICE} c) we need to condition on $A$, $L_0$ and $L_1$ to block the open pathways between $Y^{a,r=0}$ and $R$. This means accounting for all common causes of $Y$ and $R$. In this case, we have that $Y^{a,r=0} \indep R|A=a,L_0,L_1$ (\emph{conditional exchangeability}). Finally we require that there is a positive probability of \emph{not} having $R$: $P(R=0|A,L_0,L_1)>0$ for all combinations of $A,L_0,L_1$ which can occur, as we are only interested in the no-ICE (of type $R$) potential outcomes \citep{Olarte2023hypothetical}.

In many settings, assuming that the ICEs neither affect each other, nor share a common cause, may not be realistic.. For example, lack of efficacy can lead to either initiation of rescue medication and/or discontinuation of randomised treatment. Figure \ref{fig:simultaneous_ICE} b) depicts an unmeasured common cause $U$ of $D$ and $R$ that does not directly affect the outcome. The relevance of such a potential $U$ comes in the pathway from $R$ to $Y$ because it would create the biasing pathway $R \leftarrow U \rightarrow D \rightarrow Y$. As we cannot control for $U$ because it is unmeasured, we would need to adjust for $D$ to block the pathway in this setting. Given that $D$ lies on a causal pathway from $A$ to $Y$, $D$ cannot be adjusted for using standard methods. One potential solution is to use methods for time-varying confounders, accounting for $D$ in the same way as $L_1$ in the analysis \citep{Naimi2017G-methods}. Note that if the common cause $U$ of $D$ and $R$ has a direct effect on the outcome $Y$, the conditional exchangeability assumption would be violated because we cannot control for $U$ as it is unmeasured. 

\begin{figure}
    \centering
        \begin{minipage}{0.45\textwidth}
    
    a) \resizebox{0.9\textwidth}{!}{
    \begin{tikzpicture}
    \node (A) at (1.00,1.00) {$A$};
    \node (M) at (5.00,2.00) {$D$};
    \node (H) at (6.00,0.00) {$R$};
    \node (L_0) at (0.00,4.00) {$L_0$};
    \node (L_1) at (3.00,4.00) {$L_1$};
    \node (Y) at (9.00,1.00) {$Y$};
    
    \draw [->] (L_0) edge (L_1);
    \draw [->] (L_0) edge (M);
    \draw [->] (L_0) edge (H);
    \draw [->] (L_0) to[out=90,in=90] (Y);
    
    \draw [->] (A) edge (M);
    \draw [->] (A) edge (H);
    \draw [->] (A) to[out=-90,in=-90] (Y);
    \draw [->] (A) edge (L_1);
    
    \draw [->] (M) edge (H);
    
    \draw [->] (M) edge (Y);
    \draw [->] (H) edge (Y);
    
    \draw [->] (L_1) edge (M);
    \draw [->] (L_1) edge (H);
    \draw [->] (L_1) edge (Y);
    
    \end{tikzpicture}
    }
    \end{minipage}
    \hspace{0.02\textwidth}
    \begin{minipage}{0.45\textwidth}
    
    b) \resizebox{0.9\textwidth}{!}{
    \begin{tikzpicture}
    \node (A) at (1.00,1.00) {$A$};
    \node (M) at (5.00,2.00) {$D$};
    \node (H) at (6.00,0.00) {$R$};
    \node (L_0) at (0.00,4.00) {$L_0$};
    \node (L_1) at (3.00,4.00) {$L_1$};
    \node (Y) at (9.00,1.00) {$Y$};
    \node[circle,draw] (U) at (6.00,3.50) {$U$};
    
    \draw [->] (L_0) edge (L_1);
    \draw [->] (L_0) edge (M);
    \draw [->] (L_0) edge (H);
    \draw [->] (L_0) to[out=90,in=90] (Y);
    
    \draw [->] (A) edge (M);
    \draw [->] (A) edge (H);
    \draw [->] (A) to[out=-90,in=-90] (Y);
    \draw [->] (A) edge (L_1);
    
    \draw [->] (M) edge (H);
    \draw [->] (M) edge (Y);
    \draw [->] (H) edge (Y);
    
    \draw [->] (L_1) edge (M);
    \draw [->] (L_1) edge (H);
    \draw [->] (L_1) edge (Y);

    \draw [->] (U) edge (M);
    \draw [->] (U) edge (H);

    \end{tikzpicture}
    }
    \end{minipage}
    \begin{minipage}{0.45\textwidth}
    c) \resizebox{0.9\textwidth}{!}{
    \begin{tikzpicture}
    \node (A) at (1.00,1.00) {$A$};
    \node[rectangle,draw] (a) at (1.50,1.00) {$a$};
    \node (M) at (5.00,2.00) {$D^a$};
    \node (H) at (6.00,0.00) {$R^a$};
    \node[rectangle,draw] (h) at (7.00,0.00) {$r=0$};
    \node (L_0) at (0.00,4.00) {$L_0$};
    \node (L_1) at (3.00,4.00) {$L^a_1$};
    \node (Y) at (10.00,1.00) {$Y^{a,r=0}$};
    
    \draw [->] (L_0) edge (L_1);
    \draw [->] (L_0) edge (M);
    \draw [->] (L_0) edge (H);
    \draw [->] (L_0) to[out=90,in=90] (Y);
    
    \draw [->] (a) edge (M);
    \draw [->] (a) edge (H);
    \draw [->] (a) to[out=-90,in=-90] (Y);
    \draw [->] (a) edge (L_1);
    
    \draw [->] (M) edge (H);
    
    \draw [->] (M) edge (Y);
    \draw [->] (h) edge (Y);
    
    \draw [->] (L_1) edge (M);
    \draw [->] (L_1) edge (H);
    \draw [->] (L_1) edge (Y);
    
    \end{tikzpicture}
    }
    \end{minipage}
    \hspace{0.02\textwidth}
    \begin{minipage}{0.45\textwidth}
    d) \resizebox{0.9\textwidth}{!}{
    \begin{tikzpicture}
    \node (A) at (1.00,1.00) {$A$};
    \node[rectangle,draw] (a) at (1.50,1.00) {$a$};
    \node (M) at (5.00,2.00) {$D^a$};
    \node (H) at (6.00,0.00) {$R^a$};
    \node[rectangle,draw] (h) at (7.00,0.00) {$r=0$};
    \node (L_0) at (0.00,4.00) {$L_0$};
    \node (L_1) at (3.00,4.00) {$L^a_1$};
    \node (Y) at (10.00,1.00) {$Y^{a,r=0}$};
    \node[circle,draw] (U) at (6.00,3.50) {$U$};
    
    \draw [->] (L_0) edge (L_1);
    \draw [->] (L_0) edge (M);
    \draw [->] (L_0) edge (H);
    \draw [->] (L_0) to[out=90,in=90] (Y);
    
    \draw [->] (a) edge (M);
    \draw [->] (a) edge (H);
    \draw [->] (a) to[out=-90,in=-90] (Y);
    \draw [->] (a) edge (L_1);
    
    \draw [->] (M) edge (H);
    
    \draw [->] (M) edge (Y);
    \draw [->] (h) edge (Y);
    
    \draw [->] (L_1) edge (M);
    \draw [->] (L_1) edge (H);
    \draw [->] (L_1) edge (Y);

    \draw [->] (U) edge (M);
    \draw [->] (U) edge (H);

    \end{tikzpicture}
    }
    \end{minipage}
    \caption{Graphical representation using DAGs (a,b) and SWIGs (c,d) when $A=a$ and $R=0$, where treatment $A$ is assigned at random, ICE $D$ handled by treatment policy and ICE $R$ is handled by hypothetical strategy. ICE $D$ may have a causal effect on ICE $R$ and on outcome $Y$. $L_0$, $L_1$ and $U$ denote baseline variables, variables measured at $t=1$ and unmeasured variables affecting $D$ and $U$, respectively.}
\label{fig:M_causes_H}
\end{figure}
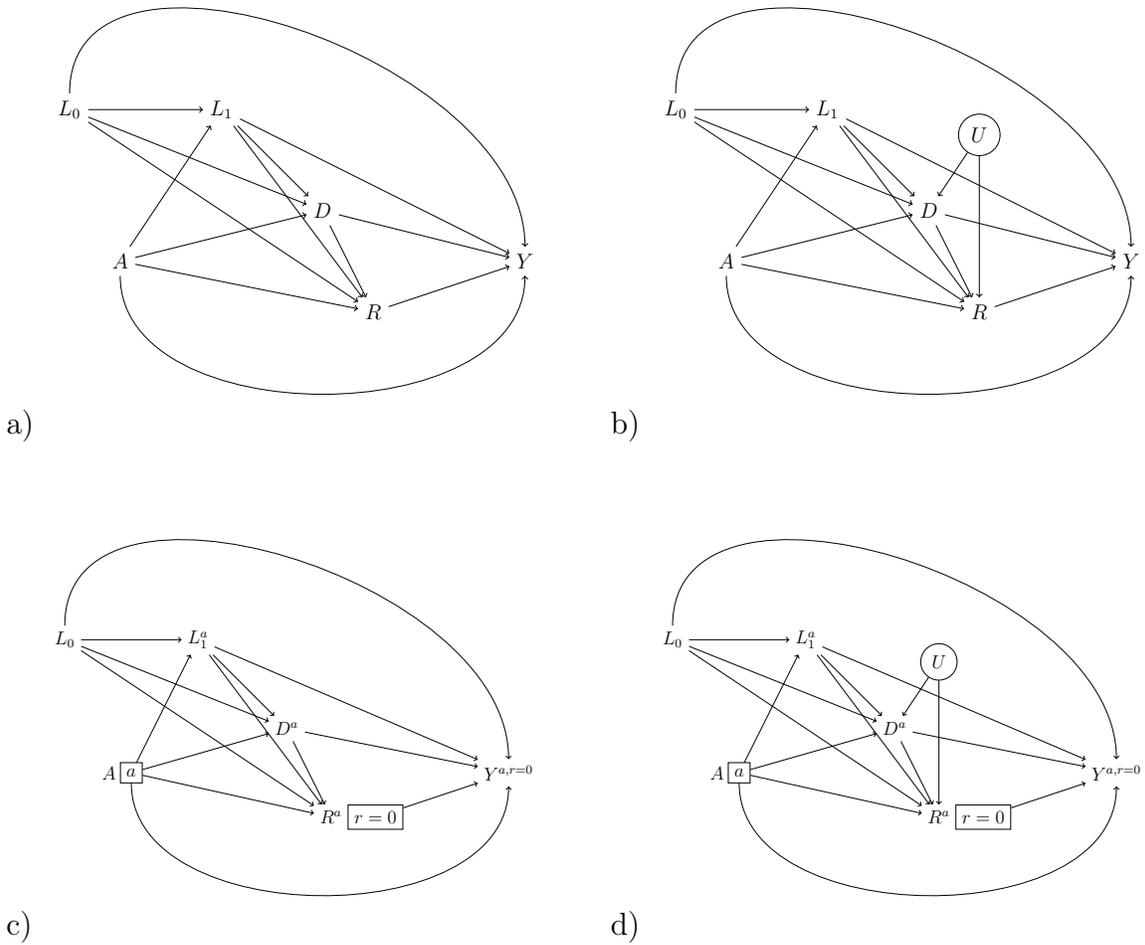

When $D$ occurs first, and has an effect on $R$ and $Y$, as shown in Figure \ref{fig:M_causes_H}, $D$ becomes a common cause of $R$ and $Y$ and plays a similar role to $L_1$. This means that the modified conditional exchangeability assumption $Y^{a,r=0} \indep R|A=a,L_0,L_1,D$ holds. 

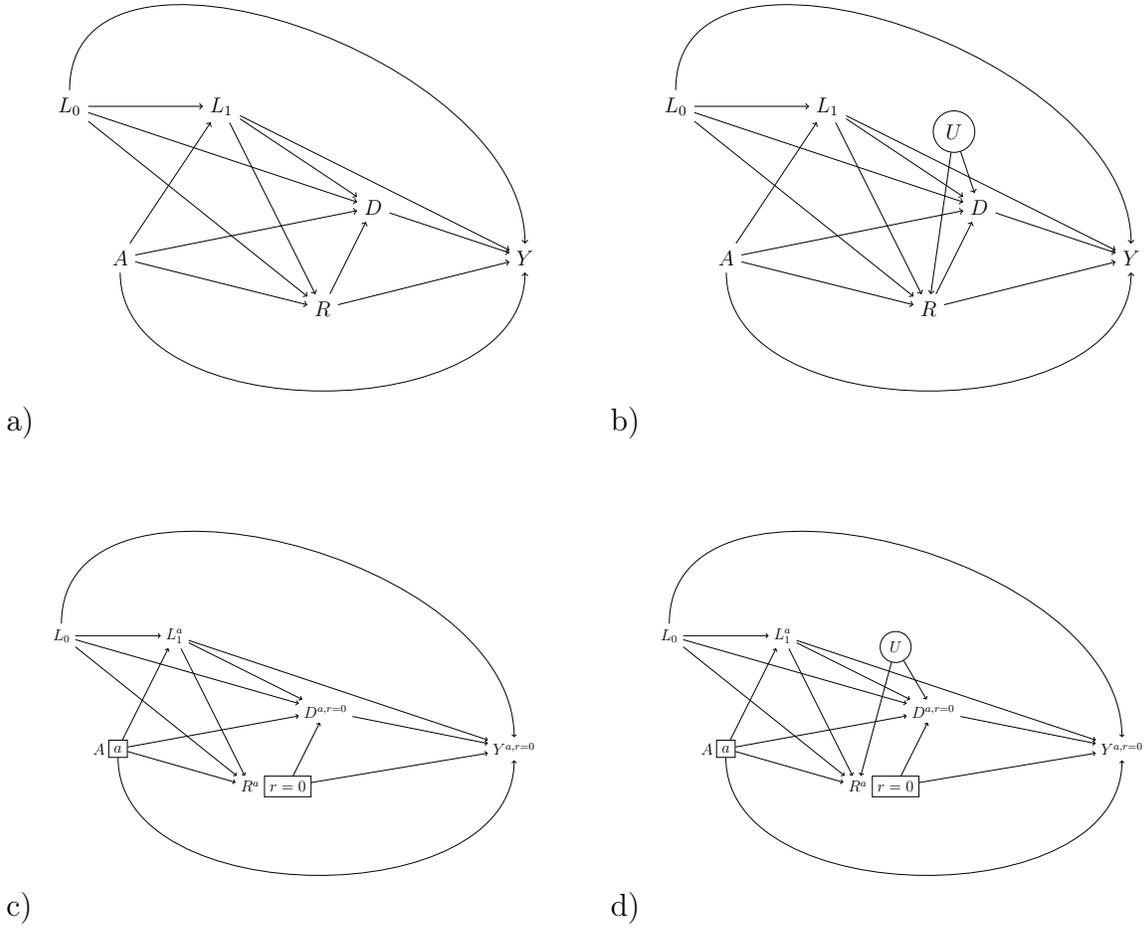
\begin{figure}
    \centering
        \begin{minipage}{0.45\textwidth}
    
    a) \resizebox{0.9\textwidth}{!}{
    \begin{tikzpicture}
    \node (A) at (1.00,1.00) {$A$};
    \node (M) at (6.00,2.00) {$D$};
    \node (H) at (5.00,0.00) {$R$};
    \node (L_0) at (0.00,4.00) {$L_0$};
    \node (L_1) at (3.00,4.00) {$L_1$};
    \node (Y) at (9.00,1.00) {$Y$};
    
    \draw [->] (L_0) edge (L_1);
    \draw [->] (L_0) edge (M);
    \draw [->] (L_0) edge (H);
    \draw [->] (L_0) to[out=90,in=90] (Y);
    
    \draw [->] (A) edge (M);
    \draw [->] (A) edge (H);
    \draw [->] (A) to[out=-90,in=-90] (Y);
    \draw [->] (A) edge (L_1);
    
    \draw [->] (H) edge (M);
    
    \draw [->] (M) edge (Y);
    \draw [->] (H) edge (Y);
    
    \draw [->] (L_1) edge (M);
    \draw [->] (L_1) edge (H);
    \draw [->] (L_1) edge (Y);
    
    \end{tikzpicture}
    }
    \end{minipage}
    \hspace{0.02\textwidth}
    \begin{minipage}{0.45\textwidth}
    
    b) \resizebox{0.9\textwidth}{!}{
    \begin{tikzpicture}
    \node (A) at (1.00,1.00) {$A$};
    \node (M) at (6.00,2.00) {$D$};
    \node (H) at (5.00,0.00) {$R$};
    \node (L_0) at (0.00,4.00) {$L_0$};
    \node (L_1) at (3.00,4.00) {$L_1$};
    \node (Y) at (9.00,1.00) {$Y$};
    \node[circle,draw] (U) at (5.50,3.50) {$U$};
    
    \draw [->] (L_0) edge (L_1);
    \draw [->] (L_0) edge (M);
    \draw [->] (L_0) edge (H);
    \draw [->] (L_0) to[out=90,in=90] (Y);
    
    \draw [->] (A) edge (M);
    \draw [->] (A) edge (H);
    \draw [->] (A) to[out=-90,in=-90] (Y);
    \draw [->] (A) edge (L_1);
    
    \draw [->] (H) edge (M);
    
    \draw [->] (M) edge (Y);
    \draw [->] (H) edge (Y);
    
    \draw [->] (L_1) edge (M);
    \draw [->] (L_1) edge (H);
    \draw [->] (L_1) edge (Y);

    \draw [->] (U) edge (M);
    \draw [->] (U) edge (H);
    
    \end{tikzpicture}
    }
    \end{minipage}
    \begin{minipage}{0.45\textwidth}
    c) \resizebox{0.9\textwidth}{!}{
    \begin{tikzpicture}
    \node (A) at (1.00,1.00) {$A$};
    \node[rectangle,draw] (a) at (1.50,1.00) {$a$};
    \node (M) at (7.00,2.00) {$D^{a,r=0}$};
    \node (H) at (5.00,0.00) {$R^a$};
    \node[rectangle,draw] (h) at (6.00,0.00) {$r=0$};
    \node (L_0) at (0.00,4.00) {$L_0$};
    \node (L_1) at (3.00,4.00) {$L^a_1$};
    \node (Y) at (12.00,1.00) {$Y^{a,r=0}$};
    
    \draw [->] (L_0) edge (L_1);
    \draw [->] (L_0) edge (M);
    \draw [->] (L_0) edge (H);
    \draw [->] (L_0) to[out=90,in=90] (Y);
    
    \draw [->] (a) edge (M);
    \draw [->] (a) edge (H);
    \draw [->] (a) to[out=-90,in=-90] (Y);
    \draw [->] (a) edge (L_1);
    
    \draw [->] (h) edge (M);
    
    \draw [->] (M) edge (Y);
    \draw [->] (h) edge (Y);
    
    \draw [->] (L_1) edge (M);
    \draw [->] (L_1) edge (H);
    \draw [->] (L_1) edge (Y);
    
    \end{tikzpicture}
    }
    \end{minipage}
    \hspace{0.02\textwidth}
    \begin{minipage}{0.45\textwidth}
    d) \resizebox{0.9\textwidth}{!}{
    \begin{tikzpicture}
    \node (A) at (1.00,1.00) {$A$};
    \node[rectangle,draw] (a) at (1.50,1.00) {$a$};
    \node (M) at (7.00,2.00) {$D^{a,r=0}$};
    \node (H) at (5.00,0.00) {$R^a$};
    \node[rectangle,draw] (h) at (6.00,0.00) {$r=0$};
    \node (L_0) at (0.00,4.00) {$L_0$};
    \node (L_1) at (3.00,4.00) {$L^a_1$};
    \node (Y) at (12.00,1.00) {$Y^{a,r=0}$};
    \node[circle,draw] (U) at (6.00,3.70) {$U$};
    
    \draw [->] (L_0) edge (L_1);
    \draw [->] (L_0) edge (M);
    \draw [->] (L_0) edge (H);
    \draw [->] (L_0) to[out=90,in=90] (Y);
    
    \draw [->] (a) edge (M);
    \draw [->] (a) edge (H);
    \draw [->] (a) to[out=-90,in=-90] (Y);
    \draw [->] (a) edge (L_1);
    
    \draw [->] (h) edge (M);
    
    \draw [->] (M) edge (Y);
    \draw [->] (h) edge (Y);
    
    \draw [->] (L_1) edge (M);
    \draw [->] (L_1) edge (H);
    \draw [->] (L_1) edge (Y);

    \draw [->] (U) edge (M);
    \draw [->] (U) edge (H);
    
    \end{tikzpicture}
    }
    \end{minipage}
    \caption{Graphical representation using DAGs (a,b) and SWIGs (c,d) of a trial when $A=a$ and $R=0$, where treatment $A$ is assigned at random, ICE $D$ handled by treatment policy and ICE $R$ is handled by hypothetical strategy. ICE $R$ may have a causal effect on $D$ and on outcome $Y$. $L_0$, $L_1$ and $U$ denote baseline variables, variables measured at $t=1$ and unmeasured variables affecting $D$ and $U$, respectively.}
    \label{fig:H_causes_M}
\end{figure}

We can also consider a scenario where ICE $R$ occurs first and can affect ICE $D$ and outcome $Y$ (Figure \ref{fig:H_causes_M}). Here $D$ lies on the causal pathways from $A$ to $Y$ and from $R$ to $Y$ and adjusting for it would introduce bias. We could consider $D$ as a time-varying covariate measured at a subsequent time point analogous to an $L_2$ and account for it using methods for time-varying confounders. Our simulations suggest however that, as one might expect, there is no  efficiency gain in doing so (Section \ref{sec:simulations}).

The cross-world hypothetical estimand (\ref{EQ:COMBINED_ESTIMAND2}) involves potential outcomes that come from two different interventional worlds, as described in Section \ref{sec:estimands}. To target such an estimand, we need to extend the assumptions discussed so far to involve both worlds. As we show in the Supplementary Material, this involves making a so-called \emph{cross-world assumption}, and under these assumptions it can be identified from the observable data.

\section{Longitudinal setting and estimation}
\label{sec:long}
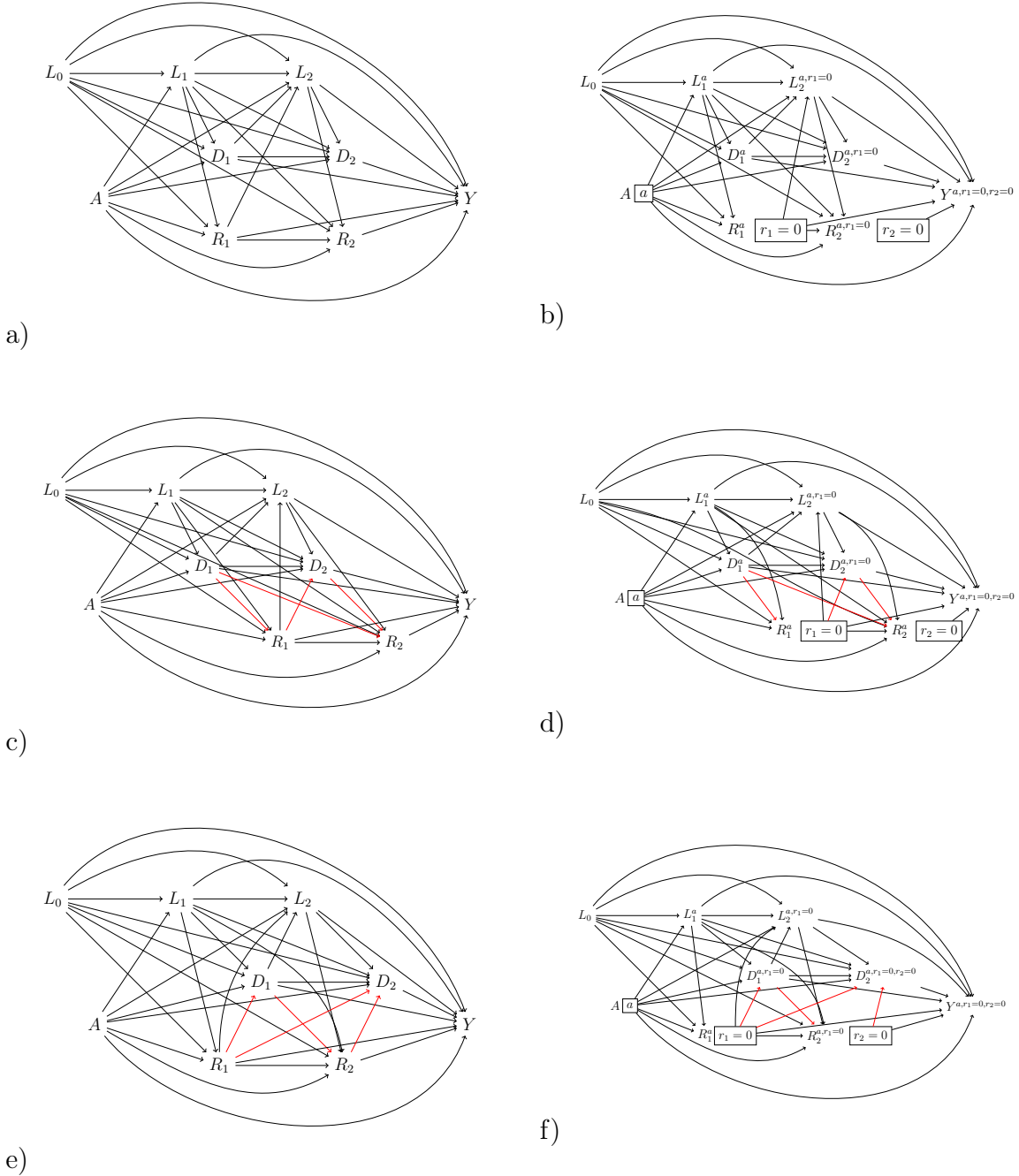
\begin{figure}
    \centering
    \begin{minipage}{0.45\textwidth}
    a) 
    \resizebox{0.9\textwidth}{!}{\begin{tikzpicture}
    \node (L_0) at (0.00,4.00) {$L_0$};
    \node (L_1) at (3.00,4.00) {$L_1$};
    \node (L_2) at (6.00,4.00) {$L_2$};
    \node (A) at (1.00,1.00) {$A$};
    \node (M_1) at (4.00,2.00) {$D_1$};
    \node (M_2) at (7.00,2.00) {$D_2$};
    \node (H_1) at (4.00,0.00) {$R_1$};
    \node (H_2) at (7.00,0.00) {$R_2$};
    \node (Y) at (10.00,1.00) {$Y$};
    
    \draw [->] (L_0) edge (L_1);
    \draw [->] (L_0) to[out=30,in=140] (L_2);
    \draw [->] (L_0) edge (M_1);
    \draw [->] (L_0) edge (M_2);
    \draw [->] (L_0) edge (H_1);
    \draw [->] (L_0) edge (H_2);
    \draw [->] (L_0) to[out=50,in=110] (Y);
    
    \draw [->] (L_1) edge (L_2);
    \draw [->] (L_1) edge (M_1);
    \draw [->] (L_1) edge (M_2);
    \draw [->] (L_1) edge (H_1);
    \draw [->] (L_1) edge (H_2);
    \draw [->] (L_1) to[out=40,in=120] (Y);
    
    \draw [->] (L_2) edge (M_2);
    \draw [->] (L_2) edge (H_2);
    \draw [->] (L_2) edge (Y);
    
    \draw [->] (A) edge (L_1);
    \draw [->] (A) edge (L_2);
    \draw [->] (A) edge (M_1);
    \draw [->] (A) edge (M_2);
    \draw [->] (A) edge (H_1);
    \draw [->] (A) to[out=-30,in=-150] (H_2);
    \draw [->] (A) to[out=-50,in=-110] (Y);

    \draw [->] (M_1) edge (L_2);
    \draw [->] (M_1) edge (M_2);
    \draw [->] (M_1) edge (Y);
    
    \draw [->] (M_2) edge (Y);

    \draw [->] (H_1) edge (L_2);
    \draw [->] (H_1) edge (H_2);
    \draw [->] (H_1) edge (Y);
    
    \draw [->] (H_2) edge (Y);
    
    \end{tikzpicture}}
    \end{minipage}
    \hspace{0.02\textwidth}
    \begin{minipage}{0.45\textwidth}
    b)
    \resizebox{0.9\textwidth}{!}{\begin{tikzpicture}
    \node (L_0) at (0.00,4.00) {$L_0$};
    \node (L_1) at (3.00,4.00) {$L^a_1$};
    \node (L_2) at (6.00,4.00) {$L^{a,r_1=0}_2$};
    \node (A) at (1.00,1.00) {$A$};
    \node[rectangle,draw] (a) at (1.50,1.00) {$a$};
    \node (M_1) at (4.00,2.00) {$D^a_1$};
    \node (M_2) at (7.20,2.00) {$D^{a,r_1=0}_2$};
    \node (H_1) at (4.00,0.00) {$R^a_1$};
    \node[rectangle,draw] (h_1) at (5.20,0.00) {$r_1=0$};
    \node (H_2) at (7.00,0.00) {$R^{a,r_1=0}_2$};
    \node[rectangle,draw] (h_2) at (8.50,0.00) {$r_2=0$};
    \node (Y) at (10.50,1.00) {$Y^{a,r_1=0,r_2=0}$};
    
    \draw [->] (L_0) edge (L_1);
    \draw [->] (L_0) to[out=30,in=140] (L_2);
    \draw [->] (L_0) edge (M_1);
    \draw [->] (L_0) edge (M_2);
    \draw [->] (L_0) edge (H_1);
    \draw [->] (L_0) edge (H_2);
    \draw [->] (L_0) to[out=50,in=110] (Y);
    
    \draw [->] (L_1) edge (L_2);
    \draw [->] (L_1) edge (M_1);
    \draw [->] (L_1) edge (M_2);
    \draw [->] (L_1) edge (H_1);
    \draw [->] (L_1) edge (H_2);
    \draw [->] (L_1) to[out=40,in=120] (Y);
    
    \draw [->] (L_2) edge (M_2);
    \draw [->] (L_2) edge (H_2);
    \draw [->] (L_2) edge (Y);
    
    \draw [->] (a) edge (L_1);
    \draw [->] (a) edge (L_2);
    \draw [->] (a) edge (M_1);
    \draw [->] (a) edge (M_2);
    \draw [->] (a) edge (H_1);
    \draw [->] (a) to[out=-30,in=-150] (H_2);
    \draw [->] (a) to[out=-50,in=-110] (Y);

    \draw [->] (M_1) edge (L_2);
    \draw [->] (M_1) edge (M_2);
    \draw [->] (M_1) edge (Y);
    
    \draw [->] (M_2) edge (Y);
    
    \draw [->] (h_1) edge (L_2);
    \draw [->] (h_1) edge (H_2);
    \draw [->] (h_1) edge (Y);
    
    \draw [->] (h_2) edge (Y);
    
    \end{tikzpicture}}
    \end{minipage}
    \begin{minipage}{0.45\textwidth}
    c)
        \resizebox{0.9\textwidth}{!}{\begin{tikzpicture}
    \node (L_0) at (0.00,4.00) {$L_0$};
    \node (L_1) at (3.00,4.00) {$L_1$};
    \node (L_2) at (6.00,4.00) {$L_2$};
    \node (A) at (1.00,1.00) {$A$};
    \node (M_1) at (4.00,2.00) {$D_1$};
    \node (M_2) at (7.00,2.00) {$D_2$};
    \node (H_1) at (6.0,0.00) {$R_1$};
    \node (H_2) at (9.00,0.00) {$R_2$};
    \node (Y) at (11.00,1.00) {$Y$};
    
    \draw [->] (L_0) edge (L_1);
    \draw [->] (L_0) to[out=30,in=140] (L_2);
    \draw [->] (L_0) edge (M_1);
    \draw [->] (L_0) edge (M_2);
    \draw [->] (L_0) edge (H_1);
    \draw [->] (L_0) edge (H_2);
    \draw [->] (L_0) to[out=50,in=110] (Y);
    
    \draw [->] (L_1) edge (L_2);
    \draw [->] (L_1) edge (M_1);
    \draw [->] (L_1) edge (M_2);
    \draw [->] (L_1) edge (H_1);
    \draw [->] (L_1) edge (H_2);
    \draw [->] (L_1) to[out=40,in=120] (Y);
    
    \draw [->] (L_2) edge (M_2);
    \draw [->] (L_2) edge (H_2);
    \draw [->] (L_2) edge (Y);
    
    \draw [->] (A) edge (L_1);
    \draw [->] (A) edge (L_2);
    \draw [->] (A) edge (M_1);
    \draw [->] (A) edge (M_2);
    \draw [->] (A) edge (H_1);
    \draw [->] (A) to[out=-30,in=-150] (H_2);
    \draw [->] (A) to[out=-50,in=-110] (Y);

    \draw [->] (M_1) edge (L_2);
    \draw [->] (M_1) edge (M_2);
    \draw [red, ->] (M_1) edge (H_1);
    \draw [red, ->] (M_1) edge (H_2);
    \draw [->] (M_1) edge (Y);

    \draw [red, ->] (M_2) edge (H_2);
    \draw [->] (M_2) edge (Y);
    
    \draw [->] (H_1) edge (L_2);
    \draw [red, ->] (H_1) edge (M_2);
    \draw [->] (H_1) edge (H_2);
    \draw [->] (H_1) edge (Y);
    
    \draw [->] (H_2) edge (Y);
    
    \end{tikzpicture}}  
    \end{minipage}
    \hspace{0.02\textwidth}
    \begin{minipage}{0.45\textwidth}
    d)
        \resizebox{0.9\textwidth}{!}{\begin{tikzpicture}
    \node (L_0) at (0.00,4.00) {$L_0$};
    \node (L_1) at (3.50,4.00) {$L^a_1$};
    \node (L_2) at (7.00,4.00) {$L^{a,r_1=0}_2$};
    \node (A) at (1.00,1.00) {$A$};
    \node[rectangle,draw] (a) at (1.50,1.00) {$a$};
    \node (M_1) at (4.50,2.00) {$D^a_1$};
    \node (M_2) at (8.00,2.00) {$D^{a,r_1=0}_2$};
    \node (H_1) at (6.00,0.00) {$R^a_1$};
    \node[rectangle,draw] (h_1) at (7.20,0.00) {$r_1=0$};
    \node (H_2) at (9.50,0.00) {$R^a_2$};
    \node[rectangle,draw] (h_2) at (10.70,0.00) {$r_2=0$};
    \node (Y) at (12.00,1.00) {$Y^{a,r_1=0,r_2=0}$};
    
    \draw [->] (L_0) edge (L_1);
    \draw [->] (L_0) to[out=30,in=140] (L_2);
    \draw [->] (L_0) edge (M_1);
    \draw [->] (L_0) edge (M_2);
    \draw [->] (L_0) edge (H_1);
    \draw [->] (L_0) to[out=-10,in=160] (H_2);
    \draw [->] (L_0) to[out=50,in=110] (Y);
    
    \draw [->] (L_1) edge (L_2);
    \draw [->] (L_1) edge (M_1);
    \draw [->] (L_1) edge (M_2);
    \draw [->] (L_1) to[out=-30,in=100] (H_1);
    \draw [->] (L_1) edge (H_2);
    \draw [->] (L_1) to[out=40,in=120] (Y);
    
    \draw [->] (L_2) edge (M_2);
    \draw [->] (L_2) to[out=-30,in=100] (H_2);
    \draw [->] (L_2) edge (Y);
    
    \draw [->] (a) edge (L_1);
    \draw [->] (a) edge (L_2);
    \draw [->] (a) edge (M_1);
    \draw [->] (a) edge (M_2);
    \draw [->] (a) edge (H_1);
    \draw [->] (a) to[out=-30,in=-150] (H_2);
    \draw [->] (a) to[out=-50,in=-110] (Y);

    \draw [->] (M_1) edge (L_2);
    \draw [->] (M_1) edge (M_2);
    \draw [red, ->] (M_1) edge (H_1);
    \draw [red, ->] (M_1) edge (H_2);
    \draw [->] (M_1) edge (Y);

    \draw [red, ->] (M_2) edge (H_2);
    \draw [->] (M_2) edge (Y);

    \draw [->] (h_1) edge (L_2);
    \draw [red, ->] (h_1) edge (M_2);
    \draw [->] (h_1) edge (H_2);
    \draw [->] (h_1) edge (Y);
    
    \draw [->] (h_2) edge (Y);
    
    \end{tikzpicture}}
    \end{minipage}
    \begin{minipage}{0.45\textwidth}
    e)
        \resizebox{0.9\textwidth}{!}{\begin{tikzpicture}
    \node (L_0) at (0.00,4.00) {$L_0$};
    \node (L_1) at (3.00,4.00) {$L_1$};
    \node (L_2) at (6.00,4.00) {$L_2$};
    \node (A) at (1.00,1.00) {$A$};
    \node (M_1) at (5.00,2.00) {$D_1$};
    \node (M_2) at (8.00,2.00) {$D_2$};
    \node (H_1) at (4.00,0.00) {$R_1$};
    \node (H_2) at (7.00,0.00) {$R_2$};
    \node (Y) at (10.00,1.00) {$Y$};
    
    \draw [->] (L_0) edge (L_1);
    \draw [->] (L_0) to[out=30,in=140] (L_2);
    \draw [->] (L_0) edge (M_1);
    \draw [->] (L_0) edge (M_2);
    \draw [->] (L_0) edge (H_1);
    \draw [->] (L_0) edge (H_2);
    \draw [->] (L_0) to[out=50,in=110] (Y);
    
    \draw [->] (L_1) edge (L_2);
    \draw [->] (L_1) edge (M_1);
    \draw [->] (L_1) edge (M_2);
    \draw [->] (L_1) edge (H_1);
    \draw [->] (L_1) to[out=-30,in=100] (H_2);
    \draw [->] (L_1) to[out=40,in=120] (Y);
    
    \draw [->] (L_2) edge (M_2);
    \draw [->] (L_2) edge (H_2);
    \draw [->] (L_2) edge (Y);
    
    \draw [->] (A) edge (L_1);
    \draw [->] (A) edge (L_2);
    \draw [->] (A) edge (M_1);
    \draw [->] (A) edge (M_2);
    \draw [->] (A) edge (H_1);
    \draw [->] (A) to[out=-30,in=-150] (H_2);
    \draw [->] (A) to[out=-50,in=-110] (Y);

    \draw [->] (M_1) edge (L_2);
    \draw [->] (M_1) edge (M_2);
    \draw [red, ->] (M_1) edge (H_2);
    \draw [->] (M_1) edge (Y);
    
    \draw [->] (M_2) edge (Y);

    \draw [->] (H_1) to[out=90,in=-140] (L_2);
    \draw [red, ->] (H_1) edge (M_1);
    \draw [red, ->] (H_1) edge (M_2);
    \draw [->] (H_1) edge (H_2);
    \draw [->] (H_1) edge (Y);
    
    \draw [red, ->] (H_2) edge (M_2);
    \draw [->] (H_2) edge (Y);
    
    \end{tikzpicture}}
    \end{minipage}
    \hspace{0.02\textwidth}
    \begin{minipage}{0.45\textwidth}
    f)
    \resizebox{0.9\textwidth}{!}{
    \begin{tikzpicture}
    \node (L_0) at (0.00,4.00) {$L_0$};
    \node (L_1) at (3.50,4.00) {$L^a_1$};
    \node (L_2) at (7.00,4.00) {$L^{a,r_1=0}_2$};
    \node (A) at (1.00,1.00) {$A$};
    \node[rectangle,draw] (a) at (1.50,1.00) {$a$};
    \node (M_1) at (6.00,2.00) {$D^{a,r_1=0}_1$};
    \node (M_2) at (10.00,2.00) {$D^{a,r_1=0,r_2=0}_2$};
    \node (H_1) at (4.00,0.00) {$R^a_1$};
    \node[rectangle,draw] (h_1) at (5.00,0.00) {$r_1=0$};
    \node (H_2) at (8.00,0.00) {$R^{a,r_1=0}_2$};
    \node[rectangle,draw] (h_2) at (9.50,0.00) {$r_2=0$};
    \node (Y) at (13.00,1.00) {$Y^{a,r_1=0,r_2=0}$};
    
    \draw [->] (L_0) edge (L_1);
    \draw [->] (L_0) to[out=30,in=140] (L_2);
    \draw [->] (L_0) edge (M_1);
    \draw [->] (L_0) edge (M_2);
    \draw [->] (L_0) edge (H_1);
    \draw [->] (L_0) edge (H_2);
    \draw [->] (L_0) to[out=50,in=110] (Y);
    
    \draw [->] (L_1) edge (L_2);
    \draw [->] (L_1) edge (M_1);
    \draw [->] (L_1) edge (M_2);
    \draw [->] (L_1) edge (H_1);
    \draw [->] (L_1) to[out=-30,in=100] (H_2);
    \draw [->] (L_1) to[out=40,in=120] (Y);
    
    \draw [->] (L_2) edge (M_2);
    \draw [->] (L_2) edge (H_2);
    \draw [->] (L_2) to[out=-10,in=130] (Y);
    
    \draw [->] (a) edge (L_1);
    \draw [->] (a) edge (L_2);
    \draw [->] (a) edge (M_1);
    \draw [->] (a) edge (M_2);
    \draw [->] (a) edge (H_1);
    \draw [->] (a) to[out=-30,in=-150] (H_2);
    \draw [->] (a) to[out=-50,in=-110] (Y);

    \draw [->] (M_1) edge (L_2);
    \draw [->] (M_1) edge (M_2);
    \draw [red, ->] (M_1) edge (H_2);
    \draw [->] (M_1) edge (Y);
    
    \draw [->] (M_2) edge (Y);

    \draw [->] (h_1) to[out=90,in=-150] (L_2);
    \draw [red, ->] (h_1) edge (M_1);
    \draw [red, ->] (h_1) edge (M_2);
    \draw [->] (h_1) edge (H_2);
    \draw [->] (h_1) edge (Y);
    
    \draw [red, ->] (h_2) edge (M_2);
    \draw [->] (h_2) edge (Y);
    
    \end{tikzpicture}
    }
    \end{minipage}
    \caption{Graphical representation using DAGs (a,c,e) and SWIGs (b,d,f) when $A=a$ and $R=0$, where treatment $A$ is assigned at random, ICE $D_k$ is handled by treatment policy and ICE $R_k$ is handled by hypothetical strategy. Both ICEs can occur at two time points. $L_0$, $L_1$ and $L_2$ denote baseline variables, variables measured at $t=1$ and at $t=2$, respectively. To emphasize whether ICE $D_k$ may have a causal effect on ICE $R_k$ or the other way round, these arrows are red.}
    \label{fig:multiple_ICE}
\end{figure}

Now we consider a setting where both types of ICE can occur at multiple time points and discuss estimation of the estimand where the ICE $D$ is handled using the treatment policy strategy and the ICE $R$ is handled using the hypothetical strategy. To keep the DAGs and SWIGs readable, we will illustrate the implications when either ICE can occur at two time points, as shown in Figure \ref{fig:multiple_ICE}, but the same idea can be extended to more time points. The first row of the figure corresponds to the situation where ICE $D_k$ and ICE $R_k$ do not affect each other throughout. But as mentioned before, this setting may be unrealistic. In the second and third rows of Figure \ref{fig:multiple_ICE}, the DAGs and SWIGs depict settings where ICE $D_k$ and ICE $R_k$ may have an effect on each other. The difference is that in the second row ICE $D_k$ precedes and affects ICE $R_k$ whereas in the third row ICE $R_k$ precedes and affects ICE $D_k$. In these cases, ICE $D_k$ becomes a common cause of ICE $R_k$ (Figure \ref{fig:multiple_ICE} c)) or $R_{k+1}$ (Figure \ref{fig:multiple_ICE} e)) and outcome $Y$. As discussed in the previous section for the setting shown in Figure \ref{fig:M_causes_H}, we then need to treat ICE variables $D_k$ as time-varying confounders, which means accounting for $D_k$ in the analysis in the same way as an $L_k$ or $L_{k+1}$ depending on the setting.

We now consider the implications for estimation of hypothetical estimand (\ref{eq:combined_estimand1}), which handles the ICE $R$ by the hypothetical strategy and the ICE $D$ by treatment policy. As described in our earlier work \citep{Olarte2023hypothetical}, a variety of estimation approaches are possible. Here we give details on estimation via multiple imputation (MI) and inverse probability weighting (IPW). To implement MI to estimate the hypothetical estimand (\ref{eq:combined_estimand1}), we can proceed by deleting values occurring after the ICE $R$ occurs and then imputing the resulting missing data. For IPW we fit models for the ICE to be handled by the hypothetical strategy. The covariates to be used in the imputation / IPW models are those required to satisfy the conditional exchangeability assumption, or equivalently, the missing at random assumption for the values made missing. For concreteness we describe how MI and IPW can be implemented to target the hypothetical estimand (\ref{eq:combined_estimand1}) under the three causal structures depicted in Figure \ref{fig:multiple_ICE} a), b) and c).

\newpage
\subsection{ICE $D$ and $R$ do not affect each other}
In this scenario (Figure \ref{fig:multiple_ICE} a)), the ICE $D$ is not a common cause of the ICE $R$ and outcome $Y$. As such it can be ignored in the estimation process. Estimation via MI can then proceed by:

\begin{enumerate}
    \item If $R_1=1$, set $L_2$ and $Y$ to missing.
    \item If $R_2=1$, set $Y$ to missing.
    \item Use MI to sequentially impute missing values in $L_2$ and $Y$:
    \begin{enumerate}
        \item Impute $L_2$ using a model with covariates $A,L_0,L_1$.
        \item Impute $Y$ using a model with covariates $A,L_0,L_1,L_2$.
    \end{enumerate}
\end{enumerate}
After imputation a regression of $Y$ on $A$ and $L_0$ can be fitted to each imputed dataset, and the estimates combined using Rubin's rules in the usual way. If desired, $D_k$ can be included, by including it in the time-varying confounders at time $k$. The resulting algorithm is then as per the MI algorithm described in the following subsection (\ref{sec:ICE_D_first}). We do not however expect including $D_k$ to confer any bias or efficiency advantage in this scenario.

Estimation via IPW involves:
\begin{enumerate}
    \item Fit regression model (e.g. logistic) for $R_1$ with covariates $A,L_0,L_1$. Calculate the (estimated) probability $\pi_1=P(R_1=0|A,L_0,L_1)$ for each patient.
    \item Fit regression model for $R_2$ with covariates $A,L_0,L_1,L_2$, using those with $R_1=0$. Calculate the probability $\pi_2=P(R_2=0|A,L_0,L_1,L_2,R_1=0)$.
\end{enumerate}
Lastly, a weighted regression of $Y$ on $A$ and $L_0$ is fitted in those with $R_1=R_2=0$, with weight equal to $(\pi_1 \pi_2)^{-1}$. The coefficient of $A$ is the estimated treatment effect for the hypothetical estimand (\ref{eq:combined_estimand1}).

\subsection{ICE $D$ precedes $R$}
\label{sec:ICE_D_first}
In this scenario (Figure \ref{fig:multiple_ICE} c)), the ICE $D$ variables are common causes of the ICE $R$ variables being handled by the hypothetical strategy and the outcome $Y$, and as such must be adjusted for in the imputation models. MI now consists of:

\begin{enumerate}
    \item If $R_1=1$, set $L_2,D_2,Y$ to missing.
    \item If $R_2=1$, set $Y$ to missing.
    \item Use MI to sequentially impute missing values in $L_2,D_2,Y$:
    \begin{enumerate}
        \item Impute $L_2$ using a model with covariates $A,L_0,L_1,D_1$.
        \item Impute $D_2$ using a model with covariates $A,L_0,L_1,D_1,L_2$.
        \item Impute $Y$ using a model with covariates $A,L_0,L_1,D_1,L_2,D_2$.
    \end{enumerate}
\end{enumerate}
Here, compared to the algorithm in the case where $D$ and $R$ do not affect each other, the ICE variable $D_k$ simply becomes part of the collection of time-varying covariates adjusted for at visit $k$.

IPW estimation consists of:
\begin{enumerate}
    \item Fit regression model (e.g. logistic) for $R_1$ with covariates $A,L_0,L_1,D_1$. Calculate the (estimated) probability $\pi_1=P(R_1=0|A,L_0,L_1,D_1)$ for each patient.
    \item Fit regression model for $R_2$ with covariates $A,L_0,L_1,L_2,D_1,D_2$, using those with $R_1=0$. Calculate the probability $\pi_2=P(R_2=0|A,L_0,L_1,L_2,D_1,D_2,R_1=0)$.
\end{enumerate}

\subsection{ICE $R$ precedes $D$}
In this scenario (Figure \ref{fig:multiple_ICE} e)), the ICE $D$ variables are again common causes of the ICE $R$ variables being handled by the hypothetical strategy and the outcome $Y$. However, because of the altered assumed temporal ordering of $D$ and $R$, $D_k$ now effectively becomes part of the time-varying confounders at visit $k+1$, rather than visit $k$. Thus MI now consists of:

\begin{enumerate}
    \item If $R_1=1$, set $D_1,L_2,D_2,Y$ to missing.
    \item If $R_2=1$, set $D_2$ and $Y$ to missing.
    \item Use MI to sequentially impute missing values in $D_1,L_2,D_2,Y$:
    \begin{enumerate}
        \item Impute $D_1$ using a model with covariates $A,L_0,L_1$.
        \item Impute $L_2$ using a model with covariates $A,L_0,L_1,D_1$.
        \item Impute $D_2$ using a model with covariates $A,L_0,L_1,D_1,L_2$.
        \item Impute $Y$ using a model with covariates $A,L_0,L_1,D_1,L_2,D_2$.
    \end{enumerate}
\end{enumerate}

IPW estimation consists of:
\begin{enumerate}
    \item Fit regression model (e.g. logistic) for $R_1$ with covariates $A,L_0,L_1$. Calculate the (estimated) probability $\pi_1=P(R_1=0|A,L_0,L_1)$ for each patient.
    \item Fit regression model for $R_2$ with covariates $A,L_0,L_1,L_2,D_1$, using those with $R_1=0$. Calculate the probability $\pi_2=P(R_2=0|A,L_0,L_1,L_2,D_1,R_1=0)$.
\end{enumerate}

As we have discussed, the assumed causal structure  determines the variables to be imputed and the order, in the case of MI, and in the case of IPW, which variables enter as covariates in the models for the ICE variables $R_k$. Table \ref{tab:summary} summarises these implications for estimating the target treatment effect when addressing an ICE with treatment policy and an ICE with hypothetical strategy simultaneously.

We emphasize that these procedures target estimand hypothetical estimand (\ref{eq:combined_estimand1}). To target the cross-world hypothetical estimand (\ref{EQ:COMBINED_ESTIMAND2}) using MI, we use all the observed values of $D$s (i.e. do not set to missing nor impute them), regardless of the occurrence of $R$s, because for this estimand the $D$s take on their natural values under the assigned treatment. 

\section{Simulations}
\label{sec:simulations}
In this section we report the results of a simulation study to demonstrate the importance of considering the causal structure when estimating the treatment effect when ICE $D$ is handled with the treatment policy strategy and ICE $R$ with a hypothetical strategy. For simplicity, we consider a setting where ICE $D_k$ and ICE $R_k$ can occur at two time points i.e. $K=2$ (Figure \ref{fig:multiple_ICE}). The R code used for the simulations is available at:  \url{https://github.com/colartep/hypothetical_and_treatment_policy}.

First, we simulated the scenario where ICE $D_k$ and ICE $R_k$ do not affect each other (Figure \ref{fig:multiple_ICE} a)). We created $10,000$ datasets for $n=2,000$ subjects as follows:
\begin{itemize}
    \item $L_0 \sim N(0, 1)$
    \item $P(A=1) = 0.5$
    \item $L_1 \sim N(\beta L_0 + \beta A, 1)$ 
    \item $P(D_1=1 | L_0, A, L_1) = \text{expit} (\alpha + \beta L_0 + \beta A + \beta L_1)$ 
    \item $P(R_1=1 | L_0, A, L_1)=\text{expit} (\alpha + \beta L_0 + \beta A + \beta L_1)$
    \item $L_2 \sim N(\beta L_0 + \beta A + \beta L_1 + \gamma D_1 + \gamma R_1, 1)$ 
    \item $P(D_2=1 | L_0, A, L_1, D_1, L_2) =\text{expit}(\alpha + \beta L_0 + \beta A + \beta L_1 + \gamma D_1 + \beta L_2)$ 
    \item $P(R_2=1 | L_0, A, L_1, R_1, L_2) = \text{expit} (\alpha + \beta L_0 + \beta A + \beta L_1 + \gamma R_1 + \beta L_2)$
    \item $Y \sim N(\beta L_0 + \beta A + \beta L_1 + \gamma D_1 + \gamma R_1 + \beta L_2 + \gamma D_2 + \gamma R_2, 1)$
\end{itemize}
where $\alpha=-1$, $\beta=0.25$, $\gamma=1$.

In the following scenario that we considered, the two ICE types affected each other, with $D_k$ preceding $R_k$ (Figure \ref{fig:multiple_ICE} c)). To simulate this setting, we made the following modifications:

\begin{itemize} 
    \item $P(R_1=1 | L_0, A, L_1, D_1)=\text{expit} (\alpha + \beta L_0 + \beta A + \beta L_1 + \gamma D_1)$
    \item $P(D_2=1 | L_0, A, L_1, D_1, R_1, L_2) =\text{expit}(\alpha + \beta L_0 + \beta A + \beta L_1 + \gamma D_1 + \gamma R_1 + \beta L_2)$ 
    \item $P(R_2=1 | L_0, A, L_1, D_1, R_1, L_2, D_2) = \text{expit} (\alpha + \beta L_0 + \beta A + \beta L_1 + \gamma D_1 + \gamma R_1 + \beta L_2 + \gamma D_2)$
\end{itemize}

Finally, we considered a scenario where ICE $R_k$ preceded $D_k$ (Figure \ref{fig:multiple_ICE} e)). For this scenario, we adapted both the way that $R_k$ and $D_k$ were generated and the order i.e. $R_k$ was simulated before $D_k$, as follows: 

\begin{itemize}
    \item $P(R_1=1 | L_0, A, L_1)=\text{expit} (\alpha + \beta L_0 + \beta A + \beta L_1)$
    \item $P(D_1=1 | L_0, A, L_1, R_1) = \text{expit} (\alpha + \beta L_0 + \beta A + \beta L_1  + \gamma R_1)$
    \item $P(R_2=1 | L_0, A, L_1, R_1, D_1, L_2) = \text{expit} (\alpha + \beta L_0 + \beta A + \beta L_1  + \gamma R_1 + \gamma D_1 + \beta L_2)$
    \item $P(D_2=1 | L_0, A, L_1, R_1, D_1, L_2, R_2) =\text{expit}(\alpha + \beta L_0 + \beta A + \beta L_1 + \gamma R_1 + \gamma D_1 + \beta L_2 + \gamma R_2)$ 
\end{itemize}

For each setting, the potential outcomes under treatment were simulated by creating a dataset of $n=10,000,000$ subjects and setting $A=1$ and $R_1=R_2=0$ in the above equations. Similarly, the potential outcome under control was estimated by creating a dataset of $n=10,000,000$ subjects and setting $A=0$ and $R_1=R_2=0$. The true effect was estimated as the difference in mean outcome $Y$ between the two groups. 

Each dataset simulated under each setting was analysed first using a `naive' estimator which regress $Y$ on $A$ and $L_0$ among those who did not experience the $R$ ICE ($R_1=R_2=0$). Next, the three variants of IPW described in Section \ref{sec:long} were used.

\begin{figure}
    \centering
    \includegraphics[width=1\linewidth]{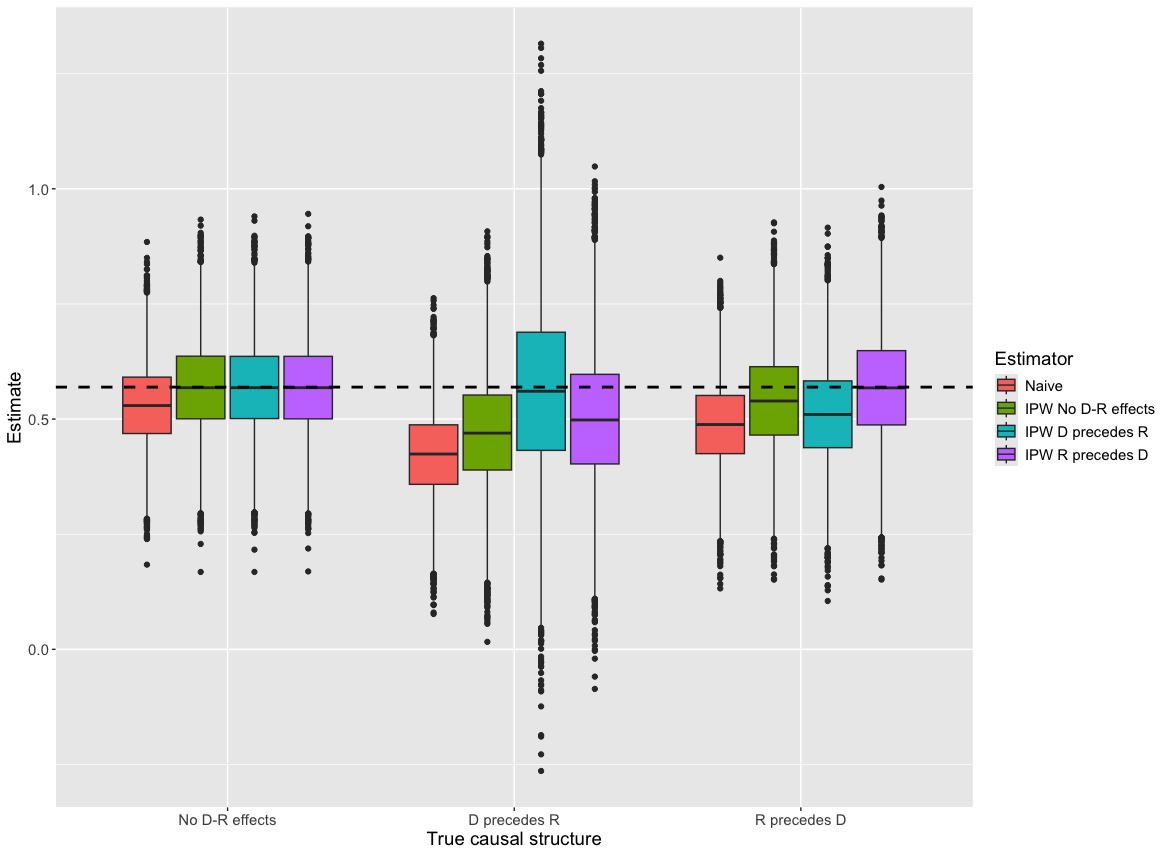}
    \caption{Simulation results where i) ICE $D$ and $R$ do not affect each other, ii) $D$ and $R$ affect each other, with $D$ preceding $R$, iii) $D$ and $R$ affect each other, with $R$ preceding $D$. Box plots show estimates (median, 25\% and 75\% centiles) from naive estimator and the three IPW estimators described in Section \ref{sec:long}.}
    \label{fig:combined-sim-results}
\end{figure}

Figure \ref{fig:combined-sim-results} shows the simulation results. The first thing to notice is that the true effect is the same across the different three true causal structures. The difference between them is how ICE $D_k$ and ICE $R_k$ were generated. As we are only interested in the $Y$ potential outcomes when $R_k=0$, then tracing the altered data generating steps under $R_k=0$ one sees that the resulting potential outcomes generated are the same across the three structures, and hence the underlying true effect is the same. 

The naive estimator is biased for the true effect in each of the three scenarios. In the first causal structure, where the two ICE types do not affect each other, all three IPW estimators are unbiased. Under this first structure, the second and third IPW estimators include additional covariates whose true coefficients are zero, such that no bias is induced. In the second scenario, where $D_k$ precedes $R_k$, only the corresponding IPW estimator is unbiased, with the other two showing some bias. This is because the `No $D$-$R$ effects' and `$R$ affects $D$' estimators omit covariates required for the conditional exchangeability assumption to hold. In the last scenario, where $R$ precedes $D$, again only the IPW estimator whose specification respects the true causal structure is unbiased.

\section{Revisiting the motivating example}
\label{sec:revisiting}
As we discussed in Section \ref{sec:example}, different stakeholders may be interested in different estimands that result from handling intercurrent events differently. Once the estimand is chosen, we have discussed that the causal order of the intercurrent events affects the identification and estimation process. We now discuss possible scenarios for the ordering of the intercurrent events in the diabetes example.

At each follow-up visit different biomarkers and diagnostic tests are recorded. In case of a (serious) adverse event, treatment discontinuation may be considered.

If this not the case, then their various biomarkers are measured, which in the diabetes example include FPG and HbA1c. When the expected levels of FPG or HbA1c are not achieved, it is reasonable to consider using rescue medication before discontinuing treatment. If the required glycaemic control is not reached even with rescue medication, then the patient may benefit from discontinuing treatment and having an alternative drug, which is decided in a following visit. 

This scenario matches the DAG in Figure \ref{fig:multiple_ICE} c) where the discontinuation node $D$ precedes the rescue node $R$ in each time point. If we follow EMEA guidelines and handle use of rescue medication with the hypothetical strategy and treatment discontinuation with treatment policy then the treatment discontinuation at time $k$, $D_k$ should be handled as part of $L_{k}$. We can proceed with estimation following the algorithm described in Section \ref{sec:ICE_D_first} that corresponds to $D$ affecting $R$.

\section{Conclusion}
\label{sec:conc}

After identifying the relevant ICEs to define the estimand of interest in a particular clinical setting, different strategies to handle them may be considered appropriate. In this paper, we have focused on having different ICEs dealt with either by the treatment policy strategy, the hypothetical strategy or both. Using potential outcome notation and causal diagrams, we discussed the different resulting estimands and whether the causal effect is identifiable or not depending on the assumed data structure. Ocampo \textit{et al}\citep{Ocampo2022SWIG} showed the relevance of using SWIGs when defining the estimand of interest in a trial, particularly because it involves potential outcomes, which are not shown in DAGs. Here, we exploited SWIGs to state the assumptions required to identify the treatment effect under different scenarios and link them with the implications for estimation. We showed that what is assumed regarding causal relationships between the ICE $D$ handled by the treatment policy strategy and the ICE $R$ handled by the hypothetical strategy has implications for if and how $D$ should be used in the statistical analysis. 

We have moreover shown that there exist at least two different causal estimands corresponding to the description `one ICE is handled using the treatment policy strategy and another with the hypothetical strategy', and that estimation must be tailored according to which is of interest. In the first, on which have focused our attention, the ICE handled using the treatment policy is (generally) affected by the intervention on the ICE handled using the hypothetical strategy. In the other, the ICE handled by treatment policy takes its natural value, i.e. the value it would take in the actual trial, even when the other ICE is intervened on by using the hypothetical strategy. We emphasized that the latter estimand does not correspond to a trial that could, even in principle, be run. As such, even though  we have demonstrated that it can be identified from the observable data under a cross-world assumption, we would generally doubt the usefulness of this estimand to stakeholders.

\begin{table}[]
    \caption{Implications for estimation when handling ICE with treatment policy and hypothetical strategy simultaneously}
    \resizebox{\textwidth}{!}{\begin{tabular}{l|l}
    Causal structure & Implication for estimation \\
    \hline
    The ICEs do not affect each other. & The treatment policy ICE can be ignored \\
    Treatment policy ICE affects hypothetical ICE & Include treatment policy ICE $D_k$ as an  $L_k$ variable  \\
    Hypothetical ICE affects treatment policy ICE & Include treatment policy ICE $D_k$ as an  $L_{k+1}$ variable  \\
    \end{tabular}}
    \label{tab:summary}
\end{table}

We considered a setting with two ICEs, but in practice often more than two will be identified at the point at which the protocol and statistical analysis plans are written. Accommodating additional ICEs that are handled by treatment policy strategy is in principle straightforward: they become part of the time-varying confounders to be adjusted for, with consideration of temporal ordering indicating to which time point's confounding variable set they belong. To handle additional ICE types by the hypothetical strategy, one approach is to combine the ICEs handled by the hypothetical strategy, i.e. the combined ICE variables at each time point become either that none of the hypothetical ICEs has yet occurred or that one or more have. Alternatively estimation methods such as inverse probability of missingness weighting can be used to model the occurrence of the different ICE types being handled by the hypothetical strategy separately. Both approaches are described in further detail in a separate paper where we analysed the diabetes trial discussed here \citep{Olarte2025diabetes}.

We have implicitly assumed, in particular in Section \ref{sec:long} where we considered estimation approaches, that there are no missing factual data (as opposed to missing counterfactual data) in the trial dataset, when in practice there often will be. Missing factual data after occurrence of ICEs being handled by the hypothetical strategy cause no issue in the estimation process. Otherwise, our general strategy, as exemplifed in our recent analysis of the diabetes trial used here as an illustrative example, is to first handle missing data via a suitable multiple imputation analysis, after which the estimand of interest can be estimated using a variety of different estimation methods \citep{Olarte2025diabetes}.

Hernan \textit{et al}\citep{Hernan2018Cautions} argue that the E9 addendum focuses too much on intercurrent events and less on treatment regimens that are clinically relevant. For instance, scenarios where treatment is \emph{not} discontinued in the presence of serious adverse event would not occur in real life and thus the treatment effect in such hypothetical cases may not be of interest. Rather than considering such hypothetical scenarios likely, they provide evidence of the added value of introducing a new drug in the market. The relevance of a particular strategy to handle each ICE depends on the specific context and should be discussed with the relevant stakeholders. Here, we discuss the different estimands both in terms of strategies to handle intercurrent events and treatment regimens that are of interest. In a given clinical setting, more than one estimand can be potentially relevant and considering all these elements could help choose between different candidates.

\section*{Data Availability Statement}
The code used to simulate the data used for this paper is available at: \url{https://github.com/colartep/hypothetical_and_treatment_policy}.

\section*{Funding Statement}
This work was funded by a UK Medical Research Council grant (MR/T023953/1 and MR/T023953/2).

\section*{Conflict of interest}
JB’s past and present institutions have received consultancy fees for his advice on statistical methodology from AstraZeneca, Bayer, Novartis, and Roche. JB has received consultancy fees from Bayer and Roche.

\bibliography{Hypothetical_and_treatment_policy}

@misc{EMAdiabetes,
  title={Guideline on clinical investigation of medicinal products in the treatment or prevention of diabetes mellitus},
  author={{Committee for Medicinal Products for Human Use}},
  howpublished={ European Medicines Society},
  year={2023}
}

@book{ICHE9Addendum,
   author = {ICH},
   title = {{International Council for Harmonisation} Topic E9(R1) on Estimands and Sensitivity Analysis in Clinical Trials},
publisher = {available at www.ich.org},
   year = {2019}
}

@article{hernan2004definition,
  title={A definition of causal effect for epidemiological research},
  author={Hern{\'a}n, Miguel Angel},
  journal={Journal of Epidemiology \& Community Health},
  volume={58},
  number={4},
  pages={265--271},
  year={2004},
  publisher={BMJ Publishing Group Ltd}
}

@misc{Hernan2018Cautions,
  title={Cautions as regulators move to end exclusive reliance on intention to treat},
  author={Hern{\'a}n, Miguel A and Scharfstein, Daniel},
  journal={Annals of internal medicine},
  volume={168},
  number={7},
  pages={515--516},
  year={2018},
  publisher={American College of Physicians},
  howpublished={}
}

@inbook{Hernan2020Ch19,
  author    = "Hernan, Miguel A and Robins, James M",
  title     = "Causal Inference: What If",
  chapter   = "19 Time-varying treatments",
  publisher = "Boca Raton: Chapman \& Hall/CRC",
  year      = "2020"
}

@article{Holzhauer2015DiabetesRescueMed,
  title={Choice of estimand and analysis methods in diabetes trials with rescue medication},
  author={Holzhauer, Bj{\"o}rn and Akacha, Mouna and Bermann, Georgina},
  journal={Pharmaceutical statistics},
  volume={14},
  number={6},
  pages={433--447},
  year={2015},
  publisher={Wiley Online Library}
}

@article{Lipkovich2020Estimands,
  title={Causal inference and estimands in clinical trials},
  author={Lipkovich, Ilya and Ratitch, Bohdana and Mallinckrodt, Craig H},
  journal={Statistics in Biopharmaceutical Research},
  volume={12},
  number={1},
  pages={54--67},
  year={2020},
  publisher={Taylor \& Francis}
}

@article{Mallinckrodt2020Estimands,
  title={Aligning estimators with estimands in clinical trials: putting the ICH E9 (R1) guidelines into practice},
  author={Mallinckrodt, CH and Bell, J and Liu, G and Ratitch, B and O’Kelly, M and Lipkovich, I and Singh, P and Xu, L and Molenberghs, G},
  journal={Therapeutic Innovation \& Regulatory Science},
  volume={54},
  number={2},
  pages={353--364},
  year={2020},
  publisher={Springer}
}

@article{Muller2018Diabetes,
  title={Efficacy and safety of dapagliflozin or dapagliflozin plus saxagliptin versus glimepiride as add-on to metformin in patients with type 2 diabetes},
  author={M{\"u}ller-Wieland, Dirk and Kellerer, Monika and Cypryk, Katarzyna and Skripova, Dasa and Rohwedder, Katja and Johnsson, Eva and Garcia-Sanchez, Ricardo and Kurlyandskaya, Raisa and Sj{\"o}str{\"o}m, C David and Jacob, Stephan and others},
  journal={Diabetes, Obesity and Metabolism},
  volume={20},
  number={11},
  pages={2598--2607},
  year={2018},
  publisher={Wiley Online Library}
}

@article{Naimi2017G-methods,
  title={An introduction to g methods},
  author={Naimi, Ashley I and Cole, Stephen R and Kennedy, Edward H},
  journal={International Journal of Epidemiology},
  volume={46},
  number={2},
  pages={756--762},
  year={2017},
  publisher={Oxford University Press}
}

@article{Ocampo2022SWIG,
  title={Single-World Intervention Graphs for Defining, Identifying, and Communicating Estimands in Clinical Trials},
  author={Ocampo, Alex and Bather, Jemar R},
  journal={Statistics in Medicine},
  year={2023},
  pages = {1-11}
}

@article{Olarte2023hypothetical,
  title={Hypothetical estimands in clinical trials: a unification of causal inference and missing data methods},
  author={Olarte Parra, Camila and Daniel, Rhian M and Bartlett, Jonathan W},
  journal={Statistics in Biopharmaceutical Research},
  volume={15},
  number={2},
  pages={421--432},
  year={2023},
  publisher={Taylor \& Francis}
}

@article{Olarte2025diabetes,
  title={Estimating hypothetical estimands with causal inference and missing data estimators in a diabetes trial case study},
  author={Olarte Parra, Camila and Daniel, Rhian M and Wright, David and Bartlett, Jonathan W},
  journal={Biometrics},
  volume={81},
  number={1},
  pages={ujae167},
  year={2025},
  publisher={Oxford University Press}
}

@article{RobinsRichardson2010Alternatives,
  title={Alternative graphical causal models and the identification of direct effects},
  author={Robins, James M and Richardson, Thomas S},
  journal={Center for the Statistics and the Social Sciences, University of Washington Series. Working Paper},
  volume={100},
  pages={1--66},
  year={2010}
}

@article{Pearl2010Consistency,
  title={On the consistency rule in causal inference: axiom, definition, assumption, or theorem?},
  author={Pearl, Judea},
  journal={Epidemiology},
  volume={21},
  number={6},
  pages={872--875},
  year={2010}
}



\bibliographystyle{agsm}

\newpage

\begin{center}
{\large\bf SUPPLEMENTARY MATERIAL}
\end{center}

\section*{Identifiability of cross-world hypothetical estimand (\ref{EQ:COMBINED_ESTIMAND2})}

In this section we show that the cross-world hypothetical estimand (\ref{EQ:COMBINED_ESTIMAND2}) is identifiable in DAG \ref{fig:H_causes_M}a) when interpreted as a non-parametric structural equation model \citep{RobinsRichardson2010Alternatives}. The cross-world hypothetical estimand (\ref{EQ:COMBINED_ESTIMAND2}) involves $E\left(Y^{a,r=0,D^{a,R^a}}\right)$ for $a=0,1$.

\subsection*{Consistency and Composition Assumptions}
We make use of the following consistency and composition assumptions. In fact these follow from DAG \ref{fig:H_causes_M}a) when this DAG is interpreted as a non-parametric structural equation model\citep{Pearl2010Consistency}.
\begin{align}
&L_1^a=L_1\textnormal{\ when\ }A=a\label{c0}\\
&R^a=R\textnormal{\ when\ }A=a\label{c1}\\
&D^{a,r}=D\textnormal{\ when\ }A=a\textnormal{\ and\ }R=r\label{c2}\\
&Y^{a,r}=Y\textnormal{\ when\ }A=a\textnormal{\ and\ }R=r\label{c3}\\
&Y^{a,r,d}=Y\textnormal{\ when\ }A=a\textnormal{,\ }R=r\textnormal{\ and\ }D=d\label{c4a}\\
&D^a=D^{a,R^a}=D^{a,r}\textnormal{\ when\ }R^a=r\label{c4}\\
&Y^{a,r,D^{a,r}}=Y^{a,r}\label{c5}\\
&Y^{a,r,D^{a,r^*}}=Y^{a,r,d}\textnormal{\ when\ }D^{a,r^*}=d\label{c6}
\end{align}
where $r={0,1}$, $r^{*}={0,1}$ and $d={0,1}$.

\subsection*{Exchangeability Assumptions}
Furthermore, we make use of the following randomisation/exchangeability assumptions, also consistent with DAG \ref{fig:H_causes_M}a):
\begin{align}A&\textnormal{\ is randomised}\label{randomisedass}\\
R&\indep D^{a,r}|L_0,A=a,L_1\label{ce1}\\
R&\indep Y^{a,r,d}|L_0,A=a,L_1\label{ce2}\\
D&\indep Y^{a,r,d}|L_0,A=a,L_1,R=r\label{ce3}
\end{align}
That (\ref{ce1}) follows from DAG \ref{fig:H_causes_M}a) can be seen from the corresponding SWIG \ref{fig:H_causes_M}c) in conjunction with (\ref{c0}) and (\ref{c1}), which allow us to replace $L_1$ and $R$ in (\ref{ce1}) with $L_1^a$ and $R^a$, respectively, and then to verify conditional independence from the SWIG. Statements (\ref{ce2}) and (\ref{ce3}) follow similarly from a SWIG that additionally includes an intervention setting $D$ to $d$.

\subsection*{Some consequences of Assumptions (\ref{c0})--(\ref{ce3})}
We note the following consequences of (\ref{c0})--(\ref{ce3}):
\begin{align}
A&\indep R^{a}|L_0,L_1^a\label{sc1}\\
A&\indep D^{a,r}|L_0,L_1^a,R^a=r\label{sc1b}\\
A&\indep Y^{a,r}|L_0,L_1^a,R^a=r\label{sc2}\\
A&\indep Y^{a,r,d}|L_0,L_1^a,R^a=r,D^a=d\label{sc2b}\\
R^a&\indep Y^{a,r,d}|L_0,L_1^a\label{sc6}\\
D^{a}&\indep Y^{a,r,d}|L_0,L_1^a,R^a=r\label{sc8}
\end{align}
The first four consequences (\ref{sc1})--(\ref{sc2b}) follow from (\ref{c0}): since $A$ has no incoming arrows, and since all arrows out of $A$ are removed in any SWIG that includes an intervention setting $A$ to $a$, there can be no path from $A$ to any other node in a SWIG depicting $R^a$, $D^{a,r}$, $Y^{a,r}$ or $Y^{a,r,d}$, and thus (\ref{sc1})--(\ref{sc2b}) hold. That (\ref{sc6}) holds can be seen directly from a SWIG such as \ref{fig:H_causes_M}c) but that additionally includes an intervention setting $D$ to $d$. That $D^{a,r}\indep Y^{a,r,d}|L_0,L_1^a,R^a=r$ follows from the same SWIG, and $D^{a,r}$ can be replaced with $D^{a}$ by assumption (\ref{c4}), leading to (\ref{sc8}).

\subsection*{Cross-world Assumption}
Finally, we will rely on the following so-called \emph{cross-world} assumption:
\begin{equation}
D^{a,r^*}\indep Y^{a,r,d}|L_0,L^a_1,R^a\label{sc10}
\end{equation}
which needs to hold even when $r$ and $r^*$ are different (hence the cross-world nature of the assumption). This does not follow from the SWIGs mentioned above, but does follow from the DAG \ref{fig:H_causes_M}a) when interpreted as a non-parametric structural equation model. However, were there to be a common cause, $L_2$, say, of $D$ and $Y$ in \ref{fig:H_causes_M}a), with $L_2$ affected by $R$, then (\ref{sc10}) would be violated.

\newpage

\subsection*{Identification}
We are now ready to show that $E\left(Y^{a,r=0,D^{a,R^a}}\right)$ and hence (\ref{EQ:COMBINED_ESTIMAND2}) can be written as a function of the distribution of the observed data under assumptions (\ref{c0})--(\ref{sc10}):

\scriptsize

\begin{align*}
E\left(Y^{a,r=0,D^{a,R^a}}\right)
&=E\left\{Y^{a,r=0,D^{a,R^a}}I(R^a=0)+Y^{a,r=0,D^{a,R^a}}I(R^a=1)\right\}\nonumber
\\&=E\left\{Y^{a,r=0}I(R^a=0)+Y^{a,r=0,D^{a,r=1}}I(R^a=1)\right\}\textnormal{\ (by assumptions (\ref{c4}) and (\ref{c5}))}\nonumber
\\&=E\left\{Y^a I(R^a=0) \right\} + E\left\{Y^{a,r=0,D^{a,R^a}}I(R^a=1)\right\}\textnormal{(since $Y^{a,r=0}=Y^{a}$ if $R^a=0$)}\nonumber
\\&=E\left\{Y I(R=0)|A=a \right\} + E\left\{Y^{a,r=0,D^{a,R^a}}I(R^a=1)\right\}\textnormal{(by assumptions (\ref{c1}), (\ref{c3}), (\ref{randomisedass}))}\nonumber
\\&=P(R=0|A=a)E(Y |A=a,R=0) + E\left\{Y^{a,r=0,D^{a,R^a}}I(R^a=1)\right\}\nonumber
\\&=P(R=0|A=a)E(Y |A=a,R=0)
+E\left\{E\left(\left.Y^{a,r=0,D^{a,r=1}}\right|L_0,L^a_1,R^a=1\right)P\left(\left.R^a=1\right|L_0,L^a_1\right)\right\}\nonumber
\\&=P(R=0|A=a)E(Y |A=a,R=0)
\\&+E\left\{\sum_d E\left(\left.Y^{a,r=0,d}\right|L_0,L^a_1,R^a=1,D^{a,r=1}=d\right)
P\left(\left.D^{a,r=1}=d\right|L_0,L^a_1,R^a=1\right)
P\left(\left.R^a=1\right|L_0,L^a_1\right)\right\}\\&\hspace{10cm}\textnormal{(by assumption (\ref{c6}))}\nonumber
\\&=P(R=0|A=a)E(Y |A=a,R=0)
\\&+E\left\{\sum_d E\left(\left.Y^{a,r=0,d}\right|L_0,L^a_1,R^a=1\right)
P\left(\left.D^{a,r=1}=d\right|L_0,L^a_1,R^a=1\right)
P\left(\left.R^a=1\right|L_0,L^a_1\right)\right\}\\&\hspace{10cm}\textnormal{(by assumption (\ref{sc10}))}\nonumber
\end{align*}

\begin{align*}
\color{white}{E\left(Y^{a,r=0,D^{a,R^a}}\right)}
\\&=P(R=0|A=a)E(Y |A=a,R=0)
\\&+E\left\{\sum_d E\left(\left.Y^{a,r=0,d}\right|L_0,L^a_1,R^a=0\right)
P\left(\left.D^{a,r=1}=d\right|L_0,L^a_1,R^a=1\right)
P\left(\left.R^a=1\right|L_0,L^a_1\right)\right\}\\&\hspace{10cm}\textnormal{(by assumption (\ref{sc6}))}\nonumber
\\&=P(R=0|A=a)E(Y |A=a,R=0)
\\&+E\left\{\sum_d E\left(\left.Y^{a,r=0,d}\right|L_0,L^a_1,R^a=0,D^a=d\right)
P\left(\left.D^{a,r=1}=d\right|L_0,L^a_1,R^a=1\right)
P\left(\left.R^a=1\right|L_0,L^a_1\right)\right\}\\&\hspace{10cm}\textnormal{(by assumption (\ref{sc8}))}\nonumber
\\&=P(R=0|A=a)E(Y |A=a,R=0)
\\&+E\left\{\sum_d E\left(\left.Y^{a,r=0,d}\right|L_0,A=a,L^a_1,R^a=0,D^a=d\right)
P\left(\left.D^{a,r=1}=d\right|L_0,A=a,L^a_1,R^a=1\right)
P\left(\left.R^a=1\right|L_0,A=a,L^a_1\right)\right\}\\&\hspace{10cm}\textnormal{(by assumptions (\ref{sc1}), (\ref{sc1b}), (\ref{sc2}) and (\ref{sc2b}))}\nonumber
\\&=P(R=0|A=a)E(Y |A=a,R=0)\nonumber
\\&+E\left\{ \left.\sum_d E\left(\left.Y\right|A=a,L_0,L_1,R=0,D=d\right)
P\left(\left.D=d\right|A=a,L_0,L_1,R=1\right)
P\left(\left.R=1\right|A=a,L_0,L_1\right) \right| A=a\right\}\\&\hspace{10cm}\textnormal{(by assumptions (\ref{c0})--(\ref{c4a}))}\nonumber
\end{align*}

\normalsize

\subsection*{Estimation}
We now show how the cross-world hypothetical estimand (\ref{EQ:COMBINED_ESTIMAND2}) can be estimated based on an imputation type approach, using:
\begin{align*}
    \frac{1}{n_a} \sum_{i: A_i=a} (1-R_i) Y_i + R_i \hat{E}(Y_i \mid A_i=a,L_{0i},L_{1i},R_i=0,D_i) 
\end{align*}
where $n_a$ denotes the number randomised to treatment group  $a$ and $\hat{E}(Y_i|A_i=a,L_{0i},L_{1i},R_i=0,D_i)$ denotes a prediction of the corresponding conditional mean based on a suitable model. This estimator is a sample mean within the $A=a$ treatment group in which for those with $R=0$ we use their observed $Y$ while for those with $R=1$ we use their predicted outcome $Y$ conditional on $A=a,L_0,L_1,D$ and $R=0$. When the model for $Y$ is correctly specified this imputation is consistent for $E\left(Y^{a,r=0,D^{a,R^a}}\right)$. To see this, note we can express the estimator as

\scriptsize
\begin{align}
    \frac{1}{n_a} \sum_{i: A_i=a} (1-R_i) Y_i + R_i \left\{ (1-D_i) \hat{E}(Y_i\mid A_i=a,L_{0i},L_{1i},R_i=0,D_i=0) + D_i \hat{E}(Y_i\mid A_i=a,L_{0i},L_{1i},R_i=0,D_i=1)\right\}
    \label{eq:altEstimandImpEst}
\end{align}

\normalsize
The first term in this estimator converges to $P(R=0|A=a)E(Y|A=a,R=0)$. The second term converges to
\begin{align*}
    E\left\{R(1-D) \lambda_{a0}(L_0,L_1) \mid A=a \right\}
\end{align*}
where $\lambda_{a0}(L_0,L_1)=E(Y|A=a,L_0,L_1,R=0,D=0)$. Using the law of total expectation, we have

\scriptsize
\begin{align*}
    E\left\{R(1-D) \lambda_0(L_0,L_1) |A=a \right\} &= E\left[E\left\{ R(1-D) \lambda_{a0}(L_0,L_1) \mid A=a,L_0,L_1\right\} \mid A=a \right] \\
    &= E\left[ \lambda_{a0}(L_0,L_1) E\left\{ R(1-D)  \mid A=a,L_0,L_1\right\} \mid A=a\right] \\
    &= E\left[\lambda_{a0}(L_0,L_1) P(R=1|A=a,L_0,L_1)P(D=0|A=a,R=1,L_0,L_1) \mid A=a\right]
\end{align*}

\normalsize
A similar argument shows the third term in \eqref{eq:altEstimandImpEst} converges to 
\begin{align*}
    E\left[\lambda_{a1}(L_0,L_1) P(R=1|A=a,L_0,L_1)P(D=1|A=a,R=1,L_0,L_1) \mid A=a\right]
\end{align*}
where $\lambda_{a1}(L_0,L_1)=E(Y \mid A=a,L_0,L_1,R=0,D=1)$, and thus we conclude the imputation estimator converges in probability to $E\left(Y^{a,r=0,D^{a,R^a}}\right)$ when the model for $E(Y_i \mid A_i=a,L_{0i},L_{1i},R_i=0,D_i)$ is correctly specified.

\end{document}